\documentstyle[12pt]{article}
\newcommand{\beq}{\begin{equation}}
\newcommand{\eeq}{\end{equation}}
\newcommand{\bea}{\begin{eqnarray}}
\newcommand{\eea}{\end{eqnarray}}

\newcommand{\g}{\gamma}
\newcommand{\k}{\kappa}
\newcommand{\e}{\epsilon}

\newcommand{\reh}{\sqrt{\epsilon \hbar}}
\newcommand{\rg}{\sqrt{g}}
\newcommand{\p}{\phi}
\newcommand{\q}{\dot{q}}

\renewcommand{\d}{\delta}
\renewcommand{\l}{\lambda}

\renewcommand{\b}{\beta}
\renewcommand{\a}{\alpha}
\newcommand{\E}{{\cal E}}
\newcommand{\ER}{\sqrt{{\cal E}}}
\newcommand{\T}{{\cal T}}
\newcommand{\A}{\mbox{\AE}}
\newcommand{\G}{{\cal G}}
\renewcommand{\H}{{\cal H}}
\newcommand{\V}{{\bf V}}
\newcommand{\N}{\tilde{N}}

\newcommand{\n}{\nu}
\newcommand{\m}{\mu}
\newcommand{\r}{\rho}
\newcommand{\s}{\sigma}
\newcommand{\D}{\Delta}
\newcommand{\th}{\theta}

\newcommand{\oh}{\frac{1}{2}}

\newcommand{\non}{\nonumber}

\renewcommand{\t}{\tau}
\newcommand{\rf}[1]{(\ref{#1})}
\newcommand{\ra}{\rightarrow}
\newcommand{\bold}{\load{\normalsize}{\bf}}
\newcommand{\pa}{\partial}
\newcommand{\sm}{\sum_{n=0}^{\infty}}
\begin{document}

\addtolength{\baselineskip}{0.20\baselineskip}

\hfill NBI-HE-94-31

\hfill gr-qc/9406044

\hfill June 1994
\begin{center}

\vspace{24pt}

{ {\Large \bf Fundamental Constants and the \\ Problem of Time}}

\end{center}

\vspace{12pt}

\begin{center}
{\sl A. Carlini$~^{\diamondsuit}$}
\footnote{Email: carlini@nbivax.nbi.dk}
{\sl and J. Greensite$~^{\ast}$}
\footnote{Permanent address:  Physics and Astronomy Dept.,
San Francisco State University, San Francisco CA 94720.
Email: greensit@stars.sfsu.edu}

\vspace{12pt}

NORDITA$~^{\diamondsuit}$ and The Niels Bohr Institute$~^{\ast}$\\
Blegdamsvej 17\\
DK-2100 Copenhagen \O, Denmark\\

\vspace{12pt}

{\bf Abstract}
\end{center}

\bigskip
\bigskip
\bigskip

   We point out that for a large class of parametrized theories,
there is a constant in the constrained Hamiltonian which
drops out of the classical
equations of motion in configuration space.  Examples
include the mass of a relativistic particle in free fall, the tension of
the Nambu string, and Newton's
constant for the case of pure gravity uncoupled to matter or other fields.
In the general case, the classically irrelevant
constant is proportional to the ratio of
the kinetic and potential terms in the Hamiltonian.  It is shown that this
ratio can be reinterpreted as an {\it unconstrained} Hamiltonian,
which generates the usual classical equations of motion.  At the quantum level,
this immediately suggests a resolution of the "problem of time" in quantum
gravity.  We then make contact with a recently proposed transfer matrix
formulation of quantum gravity and discuss the semiclassical limit.
In this formulation, it is argued that a physical
state can obey a (generalized)
Poincar\'e algebra of constraints, and still be an approximate eigenstate
of 3-geometry.  Solutions of the quantum evolution equations
for certain minisuperspace examples are presented.  An implication of
our proposal is the existence of a small, inherent uncertainty in the
phenomenological value of Planck's constant.

\vfill

\newpage

\section{Introduction}

   Actions which are invariant under reparametrizations of the time
variable typically have Hamiltonian constraints of the form
$H[p,q]=0$, which can be viewed, in an initial value problem, as
constraining the initial state $\{p,q\}_0$.
In the Dirac quantization scheme, these
constraints on the conjugate variables
become constraints on the physical states, so that
instead of the usual Schrodinger evolution equation one has

\beq
    H\left[ -i{\partial \over \partial q^a},q^a \right] \Psi[q^a] = 0
\label{Dirac}
\eeq
In cases where the
constraints are parabolic, such as the parametrized non-relativistic
particle, or parametrized scalar field theory, the constraint equation
\rf{Dirac} can often be treated exactly as a Schrodinger
equation; the coordinate whose derivatives appear only to first
order is identified with time (in the non-relativistic case), or
"many-fingered time" (in the scalar field-theory case) \cite{Kuchar1}.
On the other hand, for hyperbolic constraints such as the Klein-Gordon
equation with an arbitrary background metric $g_{\m \n}$
\beq
       \left[ -g^{\m \n}{\partial^2 \over \partial x^\m \partial x^\n}
               + m^2 \right] \psi(x^\a) = 0
\label{KG}
\eeq
or the Wheeler-de Witt equation
\bea
  & &   \left\{ -\k^2 G_{ijkl}{\d^2 \over \d g_{ij} \d g_{kl} }
+ {1 \over \k^2} \sqrt{g}(-{}^3R) \right\} \Phi = 0
\non \\
  & & G_{ijkl} = {1 \over 2\sqrt{g}}[g_{ik}g_{jl} + g_{il}g_{jk} -
g_{ij}g_{kl}]
\eea
it has proven difficult to identify an appropriate evolution
parameter and a unique, positive, and conserved
probability measure.  In quantum gravity, this difficulty is known as the
"problem of time"; c.f. ref. \cite{Kuchar} and \cite{Isham},
for recent reviews.

   In this article we propose an alternative to the conventional
procedures for quantizing parametrized theories.  Our proposal begins
with the rather trivial observation that, at the classical level, there
is no observable difference between an action $S$ and the same action
multiplied by a constant.  Since, e.g., the action for a relativistic
particle, the Nambu string, and the Einstein-Hilbert action are, respectively

\bea
       S_p &=& -m \int d\t \sqrt{-g_{\m\n}{dx^\m \over d\t}{dx^\n \over d\t}}
\non \\
       S_N &=& T \int d^2\s \sqrt{-\det \left[\eta_{\m\n}
{\partial x^\m \over \partial \s^i}
{\partial x^\n \over \partial \s^j} \right] }
\non \\
       S_{EH} &=& {1 \over \k^2} \int d^4x \; \rg (-R)
\label{actions}
\eea
it follows that the mass $m$, the string tension $T$,
and Newton's constant $G_N = \k^2/16\pi$,
which appear in the Hamiltonian constraints of the
relativistic particle, string, and
pure gravity theories respectively, do {\it not}
appear in the Euler-Lagrange equations, which are the geodesic equation
\beq
        {d^2 x^\m \over ds^2} + \Gamma^\m_{\; \a \b}
{d x^\a \over ds}{d x^\b \over ds} = 0
\eeq
for the relativistic particle, the equation of motion
\beq
      \eta^{ij} {\partial^2 \over \partial \s^i \partial \s^j}
                    x^{\m}(\s^1,\s^2) = 0
\eeq
for the Nambu string (in orthonormal coordinates),
and Einstein's equations

\beq
         R_{\m \n} = 0
\eeq
for pure gravity. The mass of a particle (or the tension of a string)
can never be determined from
its trajectory in free fall, neither can Newton's constant be determined
from vacuum solutions to Einstein's equations.  In the theories described
by \rf{actions} these are classically indeterminate parameters,  which
only become relevant by introducing non-gravitational fields
and forces (i.e. changing the theories).

   In section 2, below, we generalize this observation to any
time-parametrized theory with hyperbolic constraints, in particular,
to gravity coupled to other fields.  It will be shown
that there is always a classically undetermined parameter $\E$, which is
equal to the ratio (denoted $\A$) of the kinetic and potential terms in
the constrained Hamiltonian

\beq
     -\E = \A[p,q] = {\mbox{Kinetic}[p,q] \over \mbox{Potential}[q] }
\eeq
We will also show that replacing the usual
(constrained) Hamiltonian $H[p,q]$ in
Hamilton's equations by the expression $\A[p,q]$ generates, in configuration
space (superspace), the usual classical solutions of the theory.

   Our proposal for resolving the time problem in quantum gravity is
based on this reasoning:  since $\E$ is indeterminate at the
classical level,
we see no reason that it must be regarded as a fixed (bare) parameter
at the quantum level.  Rather, let $-\E$ denote the possible eigenvalues
of the operator $\A$.  Because $\A \ne 0$ and, as shown
in section 2, the Poisson
brackets
\beq
         \partial_\t Q = \{Q,\A\}
\label{A1}
\eeq
generate dynamics at the classical level, it is proposed that $\A$
is also the appropriate evolution operator to use at the quantum
level, i.e.
\beq
         \partial_\t <Q> = -<{i\over \hbar}[Q,\A]>
\label{A2}
\eeq
The Hilbert space of physical states is spanned by stationary
states $\Phi_\E$
\beq
         \A \Phi_\E = -\E \Phi_\E
\eeq
and this time-independent Schr\"odinger equation will be recognized as
just the usual Hamiltonian constraint with a given
value of the parameter $\E$, which varies among stationary states.

   In section 3 we rederive the $\A$ evolution operator in a completely
different way, based on a transfer-matrix quantization of parametrized
theories.  This section is a review and extension of the approach suggested
recently by one of us in ref. \cite{JG}, which leads to the same picture
as that advocated in section 2.  The transfer matrix approach
has the advantage
of fixing the operator-ordering in $\A[p,q]$ and the functional
integration measure, for parametrized theories with a discrete number
of degrees of freedom.

   Section 4 is devoted to a discussion of the semiclassical
correspondence between \rf{A1} and \rf{A2}, and in particular the
relation between the evolution parameter $\t$ that appears in those
equations, and the "many-fingered time" of general relativity.  We
argue that the space of physical states includes states which are
sharply peaked (at a particular value of $\t$) around a given
3-geometry. In such states, the dispersion around
the given 3-geometry is, crudely speaking, inversely proportional to
the dispersion in $\E$.

   We also argue that a dispersion in $\E$, which is characteristic
of non-stationary states, would appear experimentally as an inherent
uncertainty in the phenemonological value of Planck's constant.
Unfortunately, we are unable to place on this uncertainty
a reliable lower bound.

   In section 5 we illustrate our formalism by solving a simple
minisuperspace model, and display the "wavepacket of the Universe"
evolving through the stages of expansion of the scale factor,
recollapse, and reexpansion.  Section 6 contains some concluding
remarks.

\section{The Kinetic/Potential Ratio}

   We consider first a time-parametrized theory with $D$ degrees of
freedom, described by the action
\bea
     S &=& \int dt (p_a {dq^a \over dt} - N {\cal H})
\non \\
     {\cal H} &=&  {1 \over 2m_0} G^{ab}p_a p_b + m_0 V(q)
\label{St}
\eea
where the "supermetric" $G_{ab}$ has Lorentzian signature
$\{-++...+\}$.  For $V(q)=$const., this action characterizes the motion
of a relativistic particle in the spacetime metric $G_{ab}$; other
choices of $V$ arise in minisuperspace models of quantum gravity.
Making the rescaling $p_a \ra p_a /\sqrt{\E}$ (an extended canonical
transformation) the action becomes

\bea
    S &=& {1 \over \sqrt{\E}} \int dt \; \left[p_a {dq^a \over dt}
- N({1 \over 2\sqrt{\E} m_0} G^{ab} p_a p_b + \sqrt{\E} m_0 V) \right]
\non \\
      &=& {1 \over \sqrt{\E}} S[m=\sqrt{\E}m_0]
\eea
It is clear from this equation that the trajectory $q^a(t)$ in
configuration space connecting two arbitrary points $q^a_0$ and $q^a_1$
cannot depend on the value of $m_0$, which is therefore classically
indeterminate.  We may also write
\beq
   S = {1 \over \sqrt{\E}} S^\E
\eeq
where
\bea
    S^\E &=& \int dt \; (p_a {dq^a \over dt} - N \H^\E)
\non \\
    \H^\E &=& {1 \over 2 \ER m_0} G^{ab} p_a p_b + \ER m_0 V
\label{SE}
\eea
Suppose $S$ is stationary for a trajectory in phase space
$q^a(t) = Q^a(t),~~p_a(t)=P_a(t),~~N(t)={\cal N}(t)$, which connects
the generalized coordinates $q_0^a$ and $q_1^a$.  It follows that
$S^\E$ is stationary for
$q^a(t) = Q^a(t),~~p_a(t)=\sqrt{\E}P_a(t),~~N(t)={\cal N}(t)$.
This means that the trajectory in configuration space $q^a(t) = Q^a(t)$
connecting $q_0^a$ and $q_1^a$ will be the same whether the
fundamental action is taken to be $S$ or $S^\E$. It is in this
sense that the parameter $\E$ is indeterminate, since trajectories
in configuration space generated by
\beq
       \H^\E = 0~~~~~~~
       {\partial q^a \over \partial t} = N
       {\partial \H^\E \over \partial p_a}~~~~~~~
       {\partial p_a \over \partial t} = -N
       {\partial \H^\E \over \partial q^a}
\label{HamEq}
\eeq
will be independent of $\E$.

   Let us rewrite the $\H^\E=0$ constraint as
\beq
   -\E = \A[p,q] \equiv { {1 \over 2m_0} G^{ab} p_a p_b \over
m_0 V}
\eeq
It is easy to show that the configuration space trajectories
generated by treating $\A[p,q]$ as though it were a Hamiltonian
\beq
       \A=-\E ~~~~~~~
       {\partial q^a \over \partial \t} =
       {\partial \A \over \partial p_a}~~~~~~~
       {\partial p_a \over \partial \t} =
       -{\partial \A \over \partial q^a}
\label{AE-Eq}
\eeq
are equivalent to the solutions generated by \rf{HamEq}.\footnote{It
should be noted that, since $\A$ is dimensionless, the evolution parameter
$\t$ in eq. \rf{AE-Eq} has units of action.}  The
first of these equations, $\A = -\E$, is just the Hamiltonian
constraint $\H^\E=0$.  The second equation is
\bea
       {\partial q^a \over \partial \t} &=&
       {\partial \A \over \partial p_a}
\non \\
       &=& {\ER \over m_0 V}{\partial \over \partial p_a} \H^\E
\label{qt}
\eea
and for the third equation we have
\bea
   {\partial p_a \over \partial \t} &=&
       -{\partial \A \over \partial q^a}
\non \\
     &=& - {1 \over m_0 V} \left[{\partial \over \partial q^a}
({1 \over 2m_0}G^{ab}p_a p_b) - { {1 \over 2m_0}G^{ab}p_a p_b \over
m_0 V} {\partial \over \partial q^a}(m_0 V) \right]
\non \\
     &=& - {1 \over m_0 V} {\partial \over \partial q^a}
\left[{1 \over 2m_0}G^{ab}p_a p_b + \E m_0 V \right]
\non \\
     &=& - {\ER \over m_0 V} {\partial \over \partial q^a} \H^\E
\label{A3}
\eea
Suppose $\overline{q}(\t),\overline{p}(\t)$ is
some particular solution of \rf{qt}, \rf{A3}, and $\A=-\E$.  Then, choosing
\beq
        N(\t) = {1 \over V[\overline{q}(\t)]}
\eeq
and rescaling $\t$ by $\ER / m_0$ to give $\t$ the conventional units of time,
we see that this is also a solution of the Hamilton's equations and
constraint \rf{HamEq}.  It follows that, in general, the
Poisson bracket evolution equation
\beq
      \partial_\t O = \{O,\A \}
\eeq
supplemented by $\A=-\E$
will generate time evolution which is equivalent, up to a time
reparametrization, to evolution generated by
\beq
       \partial_t O = \{O,N\H^\E \}
\eeq
supplemented by $\H^\E=0$.

   The extension of these remarks to general relativity is
fairly straightforward.  Let us denote the action of a
generally covariant field theory as
\bea
        S &=& \int d^4x [ p_a {\partial q^a \over \partial t} - N \H_x -
        N_i \H^i_x]
\non \\
        \H_x &=& \k^2 G^{ab} p_a p_b + \rg U(q)
\label{S0}
\eea
where the $q^a(x)$ are the set of all fields, gravitational and
non-gravitational, $G_{ab}$ is the supermetric, $N$ and $N_i$ the
lapse and shift functions, and $\H^i_x$ are the supermomenta, linear
in the canonical momenta $p_a(x)$.   It is convenient to rescale all
non-gravitational fields by an appropriate power of $\k$ so that
all $q^a(x)$, and all $G^{ab}$, are dimensionless. In the special case of pure
gravity, we would identify
\bea
        \{a=1-6\} &\leftrightarrow& \{(i,j),~i \le j \}
\non \\
             q^a(x) &\leftrightarrow& g_{ij}(x)
\non \\
             p_a(x) &\leftrightarrow&
\left\{ \begin{array}{rr}
             p^{ij}(x)~~~~~(i=j) \\
           2 p^{ij}(x)~~~~~(i<j) \\
        \end{array} \right.
\non \\
             G_{ab}(x) &\leftrightarrow&  G^{ijnm}(x)
\non \\ U &\leftrightarrow& {1 \over \k^2}(-{}^3R)
\eea
By the same rescaling $p_a \ra p_a/ \sqrt{\E}$ as before,
the action $S$ is equivalent at the classical
level to the action
\bea
        S^\E &=& \int d^4x [ p_a {\partial q^a \over \partial t} -
N \H^{\E}_x - N_i \H^i_x]
\non \\
        \H^{\E}_x &=& {\k^2 \over \ER} G^{ab} p_a p_b + \ER \rg U(q)
\label{SA}
\eea
Variation of $S^\E$ with respect to the lapse
gives the Hamiltonian constraint
\beq
         \H^\E_x = {\k^2 \over \ER} G^{ab} p_a p_b + \ER \rg U(q) = 0
\eeq
which, multiplying again by an arbitrary function $N(x)$, we write
in the form

\beq
     -\E = {\int d^3x \; N \k^2 G^{ab} p_a p_b \over \int d^3x \;
                       N \rg U(q) }
\label{E}
\eeq
valid for any $N(x)$.
   Because of the supermomenta constraints $\H^i_x=0$, we can add a term
proportional to the supermomenta to the ratio in \rf{E} without changing its
value.  Define

\bea
      \A[p,q,N,N_i] &\equiv& \int d^3x \; \left[ {N \k^2 G^{ab} p_a p_b
     \over \int d^3x' \; \rg N U(q) } +
{1\over m_P} N_i \H^i_x \right]
\non \\
          &=& {1\over m_P} \int d^3x \; \left\{ \N \k^2 G^{ab} p_a p_b
                       + N_i \H^i_x \right\}
\label{AE}
\eea
where
\beq
         \N(x) \equiv {m_P N(x)  \over \int d^3x' \; \rg N U(q) }
\eeq
and $m_P$ is an arbitrary mass parameter.
We will now show that $\A$ can itself be reinterpreted as a Hamiltonian,
in the sense that the classical orbits in configuration space obtained from
the corresponding Hamilton's equations are identical to the extrema of
the action \rf{S0}.

   The equations of motion derived from $\A$ are
\bea
      {dq^a(x) \over d\t} &=& {\d \A \over \d p_a(x)}
\non  \\
&=&  {1\over m_P} \int d^3x' \left[ \N(x') {\d \over \d p_a(x)}
(\k^2 G^{cd}p_c p_d)_{x'}
     + N_i(x'){\d \over \d p_a(x)} \H^i_{x'} \right]
\non \\
      {dp_a(x) \over d\t} &=& -{\d \A \over \d q^a(x)}
\non \\
          &=& - {1\over m_P} \left\{ \int d^3x' \left[ \N(x')
{\d \over \d q^a(x)}(\k^2 G^{cd}p_c p_d)_{x'} + N_i(x'){\d \over \d q^a(x)}
\H^i_{x'} \right] \right.
\non \\
& &  - \left. {\int N \k^2
G^{cd}p_c p_d \over \left( \int N \rg U \right)^2}
\int d^3x' m_P N(x') {\d \over \d q^a(x)} (\rg U)_{x'} \right\}
\non \\
&=& -  {1\over m_P}\int d^3x' \left[ \N(x'){\d \over \d q^a(x)}
(\k^2 G^{cd}p_c p_d)_{x'}
- \N(x') \A {\d \over \d q^a(x)}(\rg U)_{x'} \right.
\non \\
& & \left. + N_i(x') {\d \over \d q^a(x)} \H^i_{x'} \right]
\non \\
      0 &=& {\d \A \over \d N(x)}
\non \\
&=& {1 \over \int N\rg U} (\k^2 G^{ab}p_a p_b)_x -
{ \left(\int N \k^2 G^{ab} p_a p_b \right)
\over \left(\int N \rg U \right)^2} (\rg U)_x
\non \\
 0 &=& {\d \A \over \d N_i(x)} = {1\over m_P}\H^i_x
\label{ae1}
\eea
The value of $\A$ is of course a constant of this motion, denoted
$\A=-\E$, and $\tau$ is dimensionless.  From the definition of
$\H^\E_x$ in eq. \rf{SA}, and rescaling $t=\ER \tau/m_P$ (to give $t$
the dimensions of length), the equations of motion \rf{ae1} become
\bea
      {dq^a(x) \over dt} &=&
   \int d^3x' \left[  \N(x') {\d \over \d p_a(x)} \H^\E_{x'}
     + \N_i(x'){\d \over \d p_a(x)} \H^i_{x'} \right]
\non \\
      {dp_a(x) \over dt} &=&
  - \int d^3x' \left[ \N(x') {\d \over \d q^a(x)} \H^\E_{x'}
     + \N_i(x'){\d \over \d q^a(x)} \H^i_{x'} \right]
\non \\
       \H^\E_x &=& {\k^2 \over \ER} G^{ab} p_a p_b + \ER \rg U = 0
\non \\
       \H^i_x &=& 0
\label{Heqs}
\eea
where the shift function is $\N_i \equiv N_i/\ER$.
Comparison of the equations of motion \rf{Heqs} to the Hamiltonian equations
of motion that would be derived from $S^\E$ of eq. \rf{SA} shows that they
are the same equations, apart from the restriction of the lapse function
$\N$ in \rf{Heqs} to satisfy
\beq
       \int d^3x \; \rg \N U = m_P
\label{normalize}
\eeq
The trajectory in configuration space is independent of the choice of lapse;
it is also independent, as noted above, of the choice of $\E$.

   We can therefore conclude that $\A[p,q,N,N_i]$, viewed as a
Hamiltonian, generates the same
classical dynamics as the more conventional Hamiltonian form
\beq
      H = \int d^3x [N \H_x + N_i \H^i_x]
\eeq
The crucial difference between $\A$ and $H$ is that $H$ is constrained
to vanish, in the standard formulation, whereas
the value of $\A$ is {\it unconstrained}:
the constant $\E$ can take on any value on an orbit, depending only on
the initial choice of $\{p,q\}$.  Obviously, this difference is quite
important at the quantum level.  Instead of the usual
Wheeler-de Witt constraint equation
\beq
          H\Psi[q] = 0
\label{WDeq}
\eeq
we will obtain a Schr\"odinger equation
\beq
         \A \Psi[q,\t] = i \hbar \partial_\t \Psi[q,\t]
\label{Seq}
\eeq
Several questions arise immediately:

\bigskip
\begin{description}
\item[1.] If \rf{Seq} is the dynamical equation of quantum gravity, what has
become of the Poincar\'e algebra of constraints which represents the
diffeomorphism invariance of the theory?

\item[2.] What is the relationship of the Schr\"odinger equation \rf{Seq},
based on the $\A$-operator, to path-integral quantization?

\item[3.] What is the operator ordering in $\A[p,q,N,N_i]$, and what
integration measure should be used for the inner product of states?

\item[4.] The time-evolution parameter $\t$ in the Schr\"odinger equation is
only a single variable; how is it related to the "many-fingered time" of
general relativity?
\end{description}
\bigskip

    To answer the first question (the others will be dealt with in
subsequent sections), let us note that physical states $\Psi[q,\t]$
in the Schr\"odinger representation must be independent of the
functions $N$ and $N_i$.  Expanding an arbitrary $\Psi$ in stationary
states
\beq
       \Psi[q,\t] = \sum_{\E} a_\E \Phi_\E[q] e^{i\E \t/\hbar}
\label{soln}
\eeq
where
\beq
         \A \Phi_{\E}[q] = - \E \Phi_{\E}[q]
\eeq
the condition of $N_i$-independence gives
\bea
          0 &=& m_P{\d \over \d N_i(x)} \A \Phi_\E
\non \\
            &=& \H^i_x \Phi_\E
\label{shift}
\eea
while $N$-independence requires
\bea
       0 &=& {\d \over \d N(x)} \A \Phi_\E
\non \\
         &=& {1 \over \int d^3x' \rg N U} [\k^2 G^{ab} p_a p_b -
\rg U \A]_x \Phi_\E
\non \\
         &=& {\ER \over \int d^3x' \rg N U} \H^\E_x \Phi_\E
\eea
where we have {\it provisionally} taken an operator-ordering in $\A$
with momenta to the right, and also used the supermomentum constraint
\rf{shift}.
Therefore, the Hilbert space of physical states consists of
linear combinations of $\A$-eigenstates
\beq
       \Psi_{phys}[q] = \sum_{\E} a_\E \Phi_\E[q]
\label{phys}
\eeq
each of which satisfies the constraints
\bea
       \H^\E_x \Phi_\E &=&
{1 \over \ER}[\k^2 G^{ab} p_a p_b + \E \rg U]_x \Phi_\E = 0
\non \\
       \H^i_x \Phi_\E &=& 0
\label{Econstraints}
\eea
Apart from the parameter $\E$, these are the standard constraints
on physical states of quantum gravity.
As is customary in this subject, we assume that there exists
some operator-ordering such that the commutators of operators
$\H^\E_x,~\H^i_x$ close on the Poincar\'e algebra, as do the Poisson
brackets of the corresponding classical quantities.  Given this
assumption, the physical states of the form \rf{phys} satisfy, in
our formulation, the {\it generalized} Dirac constraints
\bea
\left\{ [\k^2 G^{ab} p_a p_b]_x - [\rg U]_x \A \right\}
\Psi_{phys} &=& 0
\non \\
       \H^i_x \Psi_{phys} &=& 0
\eea

   In the case of pure quantum gravity, the $\E$ parameter has a
simple interpretation.  The Hamiltonian constraint becomes
\beq
     \left\{ {\k^2 \over \ER} G_{ijkl}p^{ij}p^{kl}
+ {\ER \over \k^2} \sqrt{g}(-{}^3R) \right\} \Phi_\E = 0
\eeq
and therefore
\beq
          G_N = {\k^2 \over 16\pi \sqrt{\E} }
\label{GNewt}
\eeq
is the effective Newton's constant for the degenerate subspace of
physical states satisfying $\A \Phi = -\E \Phi$.   The space of physical
states is spanned by states $\Phi_\E$ which satisfy the standard
Hamiltonian and supermomentum constraints, but with a different value
of Newton's constant associated with each $\E$.  In general, physical
states are not eigenstates of Newton's constant.

   To summarize our proposal for {\it pure} gravity:  since Newton's constant
is indeterminate at the classical level, we see no
reason that it is necessarily
a fixed bare parameter at the quantum level.  Instead, treating $G_N$ as
a quantum number allows for a vast extension of the space of physical
states.  It is this extension which we propose to exploit, as seen in
eq. \rf{Seq}, to resolve the time problem in quantum gravity. In the
general case of gravity coupled to other fields,
it is not Newton's constant per se but rather the kinetic/potential ratio
in the constrained Hamiltonian that is classically indeterminate.
Regarding this ratio as a q-number leads again to an extension of the
space of physical states and, as we will argue further
below, to a solution of the problem of time.

\section{The Transfer-Matrix Formulation}

   We now give an alternate derivation of the Schr\"odinger equation
\rf{Seq} for parametrized systems, following the transfer-matrix formulation
proposed by one of us in ref. \cite{JG}.  The advantage of this approach,
apart from making the connection to path-integral formalism, is that
it also fixes the measure and operator-ordering, at least for systems
with a discrete number of degrees of freedom.

   The transfer matrix $\T_\e$ for non-parametrized systems, in statistical
mechanics and Euclidean quantum mechanics, is an operator which evolves states
by a time-step $\Delta t = \e$
\bea
        \psi(q',t+\e) &=& \T_\e \psi(q',t)
\non \\
        &=&  \int d^Dq \; \m(q) \exp[- S_\e(q',q)] \psi(q,t)
\label{transfer}
\eea
Denoting by $S[q_2,q_1;\D t]$ the action of a classical solution $q(t)$
running between the initial point $q_1$ at time $t$ and final $q_2$ at
time $t+\D t$, the expression $S_\e$ in Euclidean quantum mechanics is
given by the continuation of $S$ to imaginary time lapse
\beq
         S_\e(q_2,q_1) \equiv i S[q_2,q_1;i\e]/\hbar
\label{Se1}
\eeq
The measure $\m(q)$ is defined such that $\T_\e$ is an identity operator
as $\e \ra 0$.  The usual quantum-mechanical Hamiltonian is essentially
the logarithm of the transfer matrix
\beq
        H = \lim_{\e \ra 0} (-{\hbar \over \e}) \ln[\T_\e]
\eeq
and the continuum (Euclidean) path integral is defined as the limit of a
product of transfer matrices
\bea
      \psi(q',t_1) &=& \int Dq(t_0\le t<t_1) e^{-S[q(t)]/\hbar} \psi(q,t_0)
\non \\
                     &=& \lim_{\e \ra 0} \int \prod_{n=0}^{N-1}  d^Dq_n \;
\m(q_n) \exp[-\sum_{n=0}^{N-1} S_\e(q_{n+1},q_n)] \psi(q_0,t_0)
\non \\
                     &=& \lim_{\e \ra 0} \left( \T_\e \right)^{N}
\psi(q',t_0)~~~~~~~~\mbox{where}~~~ N \equiv {t_1-t_0 \over \e}
\label{pint}
\eea

   Now let us again consider parametrized theories of the form
\bea
     S &=& \int dt (p_a {dq^a \over dt} - N {\cal H})
\non \\
     {\cal H} &=&  {1 \over 2m}G^{ab}p_a p_b + m V(q)
\label{Smini}
\eea
For parametrized theories of this sort, the transfer matrix formalism
above breaks down at eq. \rf{Se1}.  The problem is that the action
$S[q_2,q_1,G_{ab}]$ of a classical trajectory connecting initial coordinates
$q_1$ and final coordinates $q_2$ depends {\it only} on those
coordinates (and choice of supermetric $G_{ab}$);
there is no additional dependence on a time lapse $\D \t$, and the values of
parameters $\tau_1$ and $\tau_2$ that happen to be associated with
the initial and final coordinates are irrelevant. A transfer matrix
$\T_\e$ based on \rf{Se1} would then be independent of $\e$; this is
one of the obstacles encountered in trying to apply directly
the path-integral approach to find the time-evolution of physical states
in parametrized theories.

   It is premature to conclude, however, from the failure
of \rf{Se1}, that the transfer matrix concept is meaningless for
parametrized theories.  Let us denote an orbit in Hilbert space
by $\psi(q,\t)$, and require that the transfer matrix $\T_\e$ evolves
states in the parameter $\t$ such that the orbit of the center of the
wave packet
\beq
       <q^a(\t)> \equiv <\psi(q,\t)|q^a|\psi(q,\t)>
\eeq
obeys an appropriate Ehrenfest principle.  We will show, following
ref. \cite{JG}, that this is achieved by replacing \rf{Se1} in the
definition of the transfer matrix by
\beq
         S_\e(q_2,q_1) = i S[q_2,q_1,G^E_{ab}]/\reh
\label{Se2}
\eeq
where the Euclidean rotation $\D t \ra i \D t$ in \rf{Se1} is
replaced by a rotation of the signature of the supermetric
\bea
             G_{ab}(q) &=& E^i_a(q) \eta_{ij} E^j_b(q)
\non \\
    G^E_{ab} &\equiv& \mbox{sign}[V(q)] E^i_a(q) \d_{ij} E^j_b(q)
\label{euclid}
\eea

   The reason for this peculiar rotation of signature is to
ensure that $S_\e(q_2,q_1)$ is real for all choices of $q_1$ and
$q_2$.  To show this, we first compute $S[q, q+\D q,G_{ab}]$, beginning
from Hamilton's equation
\beq
       \dot{q}^a = N{\partial {\cal H} \over \partial p_a}
={N \over m} G^{ab} p_b
{}~~\Rightarrow~~ p_a = {m\over N}G_{ab}\dot{q}^b
\eeq
Insert this into the constraint equation $\H = 0$
\beq
 {1 \over 2N^2}G_{ab} \dot{q}^a \dot{q}^b  +  V = 0
\eeq
and solve for the lapse
\beq
      N = [-{1 \over 2V}G_{ab} \dot{q}^a \dot{q}^b]^{1/2}
\eeq
Then
\bea
      S[q, q+\D q,G_{ab}] &=& \int^{\D t}_0 dt \; {m \over N} G_{ab}
\dot{q}^a \dot{q}^b
\non \\
           &=& -m \int^{\D t}_0 dt \; \sqrt{-2VG_{ab} \dot{q}^a \dot{q}^b}
\non \\
           &=& -m \sqrt{-2VG_{ab} \D q^a \D q^b}
\non \\
           &=&  -m \sqrt{-\G_{ab} \D q^a \D q^b}
\label{Sq}
\eea
where we define a modified supermetric
\beq
          \G_{ab} \equiv 2 V G_{ab}
\label{Gcal}
\eeq
It is clear that if the signature of $G_{ab}$ is Lorentzian,
then $S[q, q+\D q,G_{ab}]$ is not necessarily real.  However, upon
the signature rotation \rf{euclid}, the function
\bea
    S_\e(q, q+\D q) &=& m \sqrt{2VG^E_{ab} \D q^a \D q^b}/\reh
\non \\
                  &=& m \sqrt{\G^E_{ab} \D q^a \D q^b}/\reh
\label{Semini}
\eea
is strictly real-valued for any choice of $\D q^a$.

  The integration measure $\m$ is chosen to be
\beq
      \m^{-1}(q') = (\reh )^D \lim_{\e \ra 0}\int
{d^Dq \over (\reh )^D} \; \exp(-S_\e(q',q))
\label{measure}
\eeq
This choice ensures that $\T_\e \ra 1$
in the $\e \ra 0$ limit, and that the symmetries of the action
are reflected in the measure.

   The equations
\rf{Se2},\rf{euclid},\rf{measure}, when inserted into eq. \rf{transfer},
define a transfer matrix $\T_\e$ for the parametrized, signature-rotated
theory, which in general depends on $\G^E$ and $\det[\G^E]$.  The final step
is to undo the signature rotation in computing the corresponding evolution
operator $\A$
\beq
       \A = \left[\lim_{\e \ra 0} (-{\hbar \over \e})  \ln[\T_\e]
\right]_{ \left\{ \stackrel{\scriptstyle \G^E \ra \G}
{\det(\G^E) \ra |\det(\G)|}
          \right\}  }
\label{AE-op}
\eeq
and the evolution of states is given by
\beq
       i\hbar \partial_\t \psi(q,\t) = \A \psi(q,\t)
\eeq

   We can now evaluate $\T_\e$ and $\A$ for actions of the form \rf{Smini},
leading to the function $S_\e$ shown in \rf{Semini}.  Begin by introducing
Riemann normal coordinates $\xi^a$ around $q'^a = q^a - \D q^a$, which
bring $\G^E_{ab}= \d_{ab}$ at $\xi^a=0$.  In these coordinates
\beq
      S_\e(0,\xi) = m\sqrt{\d_{ab}\xi^a \xi^b + O(\xi^5)}/\reh
\eeq
The $O(\xi^5)$ terms will not contribute to $\A$ in the $\e \ra 0$
limit, and can be dropped.  The measure is then
\bea
  \m^{-1}(q') &=& (\reh )^D \lim_{\e \ra 0} \int
{d^D\xi \over (\reh )^D}  \;
\det[{\partial \D q^\m \over \partial \xi^\n}]
     \exp[-m |\xi|/\reh]
\non \\
             &=& {2 \pi^{D/2} \over \Gamma(D/2)} (D-1)! ({\reh \over
m})^D {1 \over \sqrt{\G^E(q')}} = {\s \over \sqrt{\G^E(q')}}
\eea
The operation of the transfer matrix \rf{transfer} becomes
\bea
       \psi(q',\t+\e) &=& \int {d^D \D q \over \s }
 \sqrt{\G^E(q'+ \D q)}  \exp[-m\sqrt{\G^E_{\a \b}\D q^\a \D q^\b}/\reh ]
\psi(q'+\D q,\t)
\non \\
   &=&  \int {d^D\xi \over \s } (1 - {1\over 6}{\cal R}_{\a \b} \xi^\a
\xi^\b + ...)
 \exp[-{m |\xi| \over \reh }]
\left\{ \psi(q',\t) + {\partial \psi \over \partial \xi^\m}\xi^\m \right.
\non \\
 & & \left. + \oh{\partial^2 \psi \over \partial \xi^\m \partial \xi^\n}
\xi^\m \xi^\n + O(\xi^3) \right\}
\non \\
   &=& \left[ 1 + \e \hbar {D+1 \over 2m^2} \partial^\m \partial_\m -
\e \hbar {D+1 \over 6m^2} {\cal R} + O(\e^2) \right] \psi(q'(\xi),\t)
\eea
where ${\cal R}$ is the curvature scalar formed from the modified
supermetric $\G^E_{ab}$ of eq. \rf{Gcal}, which has been rotated to
Euclidean signature.  Transforming back from Riemann
normal coordinates, we have
\beq
  \T_\e = 1 + \e\left[ {D+1 \over 2m^2} \hbar
{1 \over \sqrt{\G^E}} {\partial
\over \partial q^n} \sqrt{\G^E} \G^{Enm} {\partial \over \partial q^m}
 -  \hbar {D+1 \over 6m^2} {\cal R}\right] + O(\e^2)
\eeq
from which we extract, according to \rf{AE-op}
\beq
      \A = -  {D+1 \over 2m^2} \hbar^2  {1 \over \sqrt{|\G|}} {\partial
\over \partial q^n} \sqrt{|\G|} \G^{nm} {\partial \over \partial q^m}
 +  \hbar^2 {D+1 \over 6m^2} {\cal R}
\label{Hmini}
\eeq
where $\G \equiv \det(\G_{mn})$, and $G_{mn}$ has been rotated back to
Lorentzian signature.

    The classical quantity $\A_{cl}$ corresponding to the operator $\A$
is obtained by replacing derivatives with c-number momenta
\beq
     \A_{cl}[q^a,p_a] = \lim_{\hbar \ra 0}
     \A[q^a,-i\hbar{\partial \over \partial q^a} \ra p_a]
\eeq
which gives
\beq
      \A_{cl} = (D+1){ {1 \over 2m} G^{ab} p_a p_b \over m V}
\eeq
This quantity, apart from an unimportant overall factor of $(D+1)$,
is simply the Kinetic/Potential ratio introduced in the previous section.
The transfer matrix approach is thus a second way of obtaining the
evolution operator $\A$, and also provides a definite prescription for
the integration measure ($\m(q) \propto |det(\G)|$), and operator
ordering.

   As a further consistency check,  we note that in ordinary
Euclidean quantum mechanics we may define
\bea
      p \psi(q') &\equiv& \lim_{\e \ra 0}
\int d^Dq \; \m ~m{(q'-q) \over \e}\exp[- S_\e(q',q)] \psi(q)
\non \\
         S_\e(q',q) &=& \left[\oh m {(q'-q)^2 \over \e}
+ V(q) \e \right]/\hbar
\eea
from which it is easy to evaluate the commutator
\bea
     [q,p]\psi &=& \lim_{\e \ra 0}\int d^Dq \; \m \{q' m{(q'-q) \over \e}
- m{(q'-q) \over \e}q \}\exp[- S_\e(q',q)] \psi(q)
\non \\
               &=& \hbar \psi
\eea
The factor $\hbar$, rather than $i\hbar$, is due to the Euclidean time.
To derive the corresponding result in the parametrized theory, we begin
from Hamilton's equation
\bea
 {\partial q^a \over \partial \t} &=& {\partial \A_{cl} \over \partial p_a}
\non \\
        &=& {(D+1) \over m^2 V} G^{ab} p_b
{}~~~~\Rightarrow~~~~p_a={m^2 V \over D+1}G_{ab}{\partial q^b \over \partial
\t}
\eea
and define
\beq
      p_a \psi(q') \equiv \lim_{\e \ra 0}
\int d^Dq \; \m(q) {m^2 V \over (D+1)} G^E_{ab}{(q'^b - q^b) \over \e}
\exp[- S_\e(q',q)] \psi(q)
\eeq
with $S_\e$ and $\m(q)$ given by \rf{Semini} and \rf{measure}.
It is then easy to show that
\beq
       [q^a,p_b] \psi = \hbar \d^a_b\psi
\eeq
as in the non-parametrized case.

   Next we consider general covariant field theories of the
form \rf{S0}.  The main obstacle to
computing  $S[q',q,G_{ab}]$, and hence to computing
the transfer matrix, is the presence of the shift functions $N_i$ in the
classical action.  Covariant derivatives of the shift functions appear in
Hamilton's equation, e.g. in the case of pure gravity
\beq
   {\partial g_{ij} \over \partial t} = 2\k^2 NG_{ijnm}p^{nm} + N_{i;j}
+ N_{j;i}
\label{p}
\eeq
and these
make it impossible to solve for the lapse-shift functions algebraically,
in terms of $\partial_t g_{ij}$.  To determine the lapse-shift
functions, it is necessary to solve certain intractable
partial differential equations.  Instead, we adopt the
strategy of simply setting $N_i=0$.  In that case, of course, the
supermomentum constraints are not obtained by extremizing the action,
and must be recovered by imposing operator constraints of the form
\beq
        Q_x[q^a,p_b] \Psi = 0
\eeq
on the physical states. These constraints (up to operator-ordering
ambiguities) will be determined below.  With the choice $N_i=0$,
Hamilton's equations give
\beq
      p_a = {1 \over 2 \k^2 N} G_{ab} \q^b
\eeq
Inserting this into the constraint equation
\beq
       0 = {\cal H}_x = {1 \over 4\k^2 N^2}G_{ab}\q^a \q^b
+ \rg U
\eeq
and solving for the lapse, gives
\beq
   N = \left[-{1 \over 4 \k^2 \rg U}G_{ab}\q^a \q^b \right]^{1/2}
\eeq
so we have
\bea
       \D S = S[q',q,G_{ab}] &=&-{1 \over \k}\int d^3x
\int^{\D t}_0 dt \sqrt{-g^{\oh} U G_{ab} \q^a \q^b}
\non \\
         &=& -{1 \over \k} \int d^3x [\sqrt{-(g^{\oh} U G_{ab})_0
\D q^a \D q^b}
                   + O(\D q^2)]
\non \\
         &=& -{1 \over \k}\int d^3x (\rg)_0 [\sqrt{-(\G_{ab})_0
\D q^a \D q^b} + O(\D q^2)]
\label{DS}
\eea
where $\D q^a = q^a-q'^a$, and we define
\beq
       \G_{ab} \equiv {1 \over \rg} U G_{ab}
\eeq
The notation $(..)_0$ means that the quantity in
parenthesis is to be evaluated at $\D q =0$; i.e. at $q'$.

   Applying the prescriptions \rf{transfer},\rf{Se2},\rf{euclid}, the
transfer matrix is formally obtained from
\bea
     \psi(q',\t+\e) &=& \int Dq \; \m(q) e^{-\D S/\reh } \left[ \psi(q')
+ \int d^3x \left({\d \psi \over \d q^a(x)}\right) \D q^a(x)
\right.
\non \\
    & & \left. + \oh \int d^3x d^3y \left( {\d^2 \psi \over \d q^a(x)
\d q^b(y)}\right) \D q^a(x) \D q^b(y) + ... \right]
\non \\
    &=& \psi(q',\t) + \e [T_0 + T_1 + T_2] + O(\e^2)
\non \\
    &=& \T_\e \psi(q',\t)
\label{Tn}
\eea
where the $T_n$ represent terms with $n$ functional derivatives of
$\psi$.  To find these terms, we need to evaluate
\bea
    <\D q^a(x_1) \D q^b(x_2)> &=& \int D(\D q) \; (\m)_0
\D q^a(x_1) \D q^b(x_2)
\non \\
& & \times \exp\left[-{1 \over \k}\int d^3x (\rg)_0 \sqrt{(\G^E_{ab})_0
\D q^a \D q^b}/\reh \right]
\label{singular}
\eea
Unfortunately this quantity is highly singular, and in fact ill-defined
without a regularization procedure of some kind.

   The authors are not aware of a non-perturbative regulator of the
integral over three-metrics in \rf{Tn} which preserves an exact diffeomorphism
invariance.  In the absence of such a regulator (which is also crucial
for sorting out the operator-ordering issues
\cite{Woodard}), we can only make some
general remarks about the regularized form of $<\D q^a(x_1) \D q^b(x_2)>$.
Inspection of \rf{singular} shows that $<\D q^a(x_1) \D q^b(x_2)>=0$
for $x_1 \ne x_2$; we also see that $(\G^E_{ab})_0 \D q^a(x) \D q^b(x)$
transforms like a scalar.  One therefore expects the regulated
expectation value to go like
\beq
      <\D q^a(x_1) \D q^b(x_2)>_{reg} \approx {\e \hbar \k^2 \over v}
(\G^{Eab} {1 \over \rg})_0 \d^3(x_1-x_2)
\label{regular}
\eeq
where $v$ is a scalar quantity with dimensions of volume, $v \ra 0$
as the regulator is removed.  A very important issue is whether $v$ has
some dependence on the 3-metric $(g_{ij})_0$, and this depends on the
properties of the (unknown) regulator.  If, as is the case with
dynamical triangulation, there is a fixed short-distance cutoff
$l$, then we expect that $v \approx l^3$, and the number of
degrees of freedom changes with the volume of the manifold.  If,
on the other hand, the number of degrees of freedom $N_p$ is
fixed (as in the Regge lattice), then the volume per degree of freedom
changes with the volume of the manifold.  In the latter case, it is
reasonable to expect that
\beq
      <\D q^a(x_1) \D q^b(x_2)>_{reg} \approx \e \hbar \k^2 {\b \over \V}
(\G^{Eab} {1 \over \rg})_0 \d^3(x_1-x_2)
\label{unleaded}
\eeq
where $\V$ is the volume of the 3-manifold described by $(g_{ij})_0$,
and $\b$ is proportional to the number of degrees of freedom in
the regularization.  A naive lattice regularization, replacing, e.g.
\bea
        \D q^a(x) &\leftrightarrow& \D q^a(n)
\non \\
        \int d^3x \rg &\leftrightarrow& {\V \over N_p} \sum_{n=1}^{N_p}
\non \\
         {\d \over \d q^a(x)}  &\leftrightarrow& {N_p \sqrt{g(n)} \over \V}
{\partial \over \partial q^a(n)}
\non \\
          Dq &\leftrightarrow& \prod_n d^Dq(n)
\label{discrete}
\eea
does, in fact, lead to \rf{unleaded} (c.f. \cite{JG}), although
of course such a regularization does not at all respect diffeomorphism
invariance.  We will assume that there exists {\it some}
good regularization leading to \rf{unleaded}, although our justification
for this assumption is largely a posteriori.

   Using eq. \rf{unleaded}, the $T_2$ term (in \rf{Tn}) contributing
to the transfer matrix is easily evaluated; it is the only term which
is important in computing the semiclassical limit.  The other terms
are operator-ordering contributions which, in the absence of an
explicit regulator, we will ignore.  The corresponding $\A$ operator
is
\beq
      "\A" = -\hbar^2 {\b \k^2 \over \V} \int d^3x \; U^{-1} G^{ab}
{\d^2 \over \d q^a \d q^b}
\label{Hgrav}
\eeq
where $"\A"$ is in quotes to emphasize that
this operator, by itself, does {\it not} yield the correct
equations of motion in the classical limit.  This was to be expected,
because $\A$ has been derived by setting the shift functions $N_i=0$,
a step which requires additional constraints on the physical states.
Now note that because of the spatial volume denominator in \rf{Hgrav},
we can write the evolution equation \rf{Seq} as
\beq
      \int d^3x Q_x \Psi = 0
\label{QS_eq}
\eeq
where
\beq
       Q_x = -\hbar^2 \k^2 U^{-1} G^{ab} {\d^2 \over \d q^a \d q^b}
 -i \hbar \rg {\partial \over \partial \t}
\eeq
and where we have absorbed the constant $\b$ in $"\A"$ into a rescaling of
the evolution parameter $\t$ in eq. \rf{Seq}.
The extra constraints which need to be imposed on the physical states,
which then generate the usual constraint algebra of general relativity,
are simply
\beq
        Q_x \Psi = 0
\label{Qconstraint}
\eeq
at every point $x$.  To show this, consider an arbitrary solution of
the evolution equation
\beq
       \Psi(q,\t) = \sum_{\E} a_\E e^{i\E \t /\hbar} \Phi_\E(q)
\eeq
Since the $a_\E$ are arbitrary, the Q-constraint \rf{Qconstraint}
requires that for each stationary state
\beq
\H^\E_x \Phi_\E \equiv
\left\{ -\hbar^2 {\k^2 \over \ER} G^{ab} {\d^2 \over \d q^a \d q^b}
+  \ER \rg U \right\} \Phi_{\E} = 0
\label{WD2}
\eeq
But this is simply the Wheeler-de Witt equation for the action
$S_\E$ of eq. \rf{SA}!  Moreover, the constraints $\H^\E_x \Phi_\E = 0$
imply, via the Moncrief-Teitelboim interconnection theorem
\cite{Moncrief}, that the supermomentum constraints
\beq
       \H^i_x \Phi_\E = 0
\label{Hi_constraint}
\eeq
are satisfied as well.  In this way, we find that the
stationary states $\{\Phi_\E\}$ satisfy the usual constraint algebra
of general relativity, given by the action $S_\E$.  The Hilbert space
of all physical states is spanned by the stationary states, with all
possible values of $\E$.

    Now multiplying both sides of \rf{Qconstraint} by $NU$, where $N$
is an arbitrary function, integrating over space, and applying
the supermomentum constraint \rf{Hi_constraint}, we have in place of
\rf{Hgrav}
\bea
   i\hbar \partial_\t \Psi &=& \left[ {1 \over \int d^3x \rg N U}
 \k^2 \int d^3x \;   N G^{ab} (-\hbar^2{\d^2 \over \d q^a \d q^b})
\right] \Psi
\non \\
   &=& {1 \over m_P} \int d^3x \; \left[ -\hbar^2  \N \k^2 G^{ab}
{\d^2 \over \d q^a \d q^b} + N_i \H^i_x \right] \Psi
\non \\
   &=& \A \Psi
\label{AE2}
\eea
where
\beq
   \A = {1 \over m_P} \int d^3x \; \left[ -\hbar^2  \N \k^2 G^{ab}
{\d^2 \over \d q^a \d q^b} + N_i \H^i_x \right]
\eeq
is the operator form of the Kinetic/Potential ratio \rf{AE}
introduced in the previous section, with
\beq
         \N(x) \equiv  m_P {N(x)  \over \int d^3x' \; \rg N U(q) }
\eeq
It should be remembered that certain operator-ordering contributions
to \rf{AE2}, coming from the $T_0$ and $T_1$ terms in \rf{Tn} have
been dropped. However, operator ordering terms are always
$O(\hbar)$, and will not affect the correspondence
of the $\A$ operator to the Kinetic/Potential ratio \rf{AE} in the
classical limit.  From the fact that
physical states are independent of $N$ and $N_i$
\beq
      {\d \over \d N(x)} \A \Psi = {\d \over \d N_i(x)} \A \Psi = 0
\eeq
we find, as in the previous section, the generalized constraints
\bea
\left\{ [\k^2 G^{ab} p_a p_b]_x - [\rg U]_x \A \right\}
\Psi_{phys} &=& 0
\non \\
       \H^i_x \Psi_{phys} &=& 0
\label{gencon}
\eea
The closure of these generalized constraints under commutation
depends only on the assumed closure of the
standard constraints \rf{Econstraints}, for any fixed value of the
parameter $\E$.

   From the Schr\"odinger evolution equation \rf{Seq}, we have
\bea
       \partial_\t <q^a> &=& {i\over \hbar}<[\A,q^a]>
     = <{\partial \A \over \partial p_a}>
{}~~+~~ \mbox{op.-ordering terms}
\non \\
&=&{1\over m_P} <\int d^3x' \left[ \N(x') {\d \over \d p_a(x)}
(\k^2 G^{cd}p_c p_d)_{x'}
     + N_i(x'){\d \over \d p_a(x)} \H^i_{x'} \right]>
\non \\ & & ~~+~~ \mbox{op.-ordering terms}
\non \\
       \partial_\t <p_a> &=& {i\over \hbar}<[\A,p_a]>
     = <-{\partial \A \over \partial q^a}> ~~+~~ \mbox{op.-ordering terms}
\non \\
&=&- <{1\over m_P} \int d^3x' \biggl[ \N (x'){\d \over \d q^a(x)}
(\k^2 G^{cd}p_c p_d)_{x'}
\non \\
& &- \N (x') \A {\d \over \d q^a(x)}(\rg U)_{x'}
+ N_i(x') {\d \over \d q^a(x)} \H^i_{x'} \biggr]>
\non \\
& & ~~+~~ \mbox{op.-ordering terms}
\label{Efest}
\eea

    Equation \rf{Efest} is the Ehrenfest principle obtained from
our transfer matrix formalism.  Removing the "$<>$" brackets,
replacing $\A$ by a constant parameter $-\E$, and
dropping operator-ordering terms, eq. \rf{gencon} and
\rf{Efest} become the classical equations of motion \rf{Heqs}.

\section{WKB, Time, and Many-Fingered Time}

   In classical general relativity, the split of spacetime into
space + time can be accomplished in infinitely many ways, which are
distinguished by a choice of lapse and shift functions.  The
geometrical meaning of a time parameter $t$, in the classical equations
of motion
\bea
  \partial_t Q &=& \{Q,H\}
\non \\
                &=& \int d^3x \left[
N(x) \{Q,\H_x\} + N_i(x) \{Q,\H_x^i\} \right]
\label{tradition}
\eea
is specified by $N(x)$ and $N_i(x)$.  This is also the case for
the $\A$-evolution equation
\bea
      \partial_\t Q &=& \{Q,\A(p,q,N,N_i)\}
\non \\
         &=& {1\over m_P}\int d^3x \left[
\ER \N(x) \{Q,\H^\E_x\} + N_i(x) \{Q,\H_x^i\} \right]
\label{c-bracket}
\eea
except that in the latter case the lapse $\N$ has been normalized
to satisfy the condition \rf{normalize}.  To see what
this restriction means for the evolution parameter $\t$, let us consider
Hamilton's principal function $S^\E_{HP}[q,q']$ satisfying the
Hamilton-Jacobi equations and constraints
\bea
        \k^2 G_{ab} {\d S^\E_{HP} \over \d q^a}{\d S^\E_{HP} \over \d q^b}
            + \E \rg U &=& 0
\non \\
         \H^i\left[ p_a={\d S^\E_{HP} \over \d q^a}\right] &=& 0
\label{HJeq}
\eea
and
\bea
    \k^2 G_{ab} {\d S^\E_{HP} \over \d q'^a}{\d S^\E_{HP} \over \d q'^b}
            + \E \rg U &=& 0
\non \\
         \H^i\left[ p'_a=-{\d S^\E_{HP} \over \d q'^a}\right] &=& 0
\label{HJcon}
\eea
As usual, the principal function $S^\E_{HP}$ defined in this way
has the interpretation of being the action of the classical trajectory
in superspace, connecting configurations $q^a(x)$ and $q'^a(x)$.
Then consider the variation of $S^\E_{HP}$ in time $\t$ according
to \rf{c-bracket}
\bea
       \D S^\E_{HP} &=& {d S^\E_{HP} \over d\t} \D \t =\{S^\E_{HP},\A \} \D \t
\non \\
 &=& \int d^3x \; {\d S^\E_{HP} \over \d q^a} {\d \A \over \d p_a} \D \t
\non \\
 &=& {1\over m_P}\int d^3x \; {\d S^\E_{HP} \over \d q^a}\left\{2\k^2 \N G^{ab}
{\d S^\E_{HP} \over \d q^b} + {\d \over \d p_a(x)} \int d^3x' N_i \H^i_{x'}
\right\} \D \t
\non \\
 &=& {1\over m_P}\int d^3x \; \left\{2\k^2 \N G^{ab}{\d S^\E_{HP} \over \d q^a}
{\d S^\E_{HP} \over \d q^b}
+ N_i \H^i_x[p_a={\d S^\E_{HP} \over \d q^a}] \right\} \D \t
\non \\
 &=& 2\A\left[p_a={\d S^\E_{HP} \over \d q^a}\right] \D \t
\non \\
 &=& -2\E \D \t
\label{DSdt}
\eea

   What eq. \rf{DSdt} establishes is that, in evolving classically from
configuration $q'$ to $q$ in time $\D \t$ according to \rf{c-bracket},
the action $S^\E_{HP}[q,q']$ is proportional to $\D \t$ {\it regardless
of the choice of} $N,N_i$.  In other words, while the particular
configuration $q$ reached after $\D \t$ does depend on $N,N_i$, the
increment of action $\D S$ does not.  This situation is indicated
schematically in Figure 1.  Of course, the restriction $\D \t \propto
\D S^\E_{HP}$ is not a restriction on hypersurfaces; $Q[q,p]$ can be
computed from \rf{c-bracket} on any spacelike hypersurface of the
classical manifold, by choosing an appropriate $N,N_i$.  For this
reason, the $\A$-evolution equation
\rf{c-bracket} is just as informative as the traditional evolution
equation \rf{tradition}.

   The quantum equation of motion for an observable Q, in our
formalism, is
\beq
      \partial_\t <Q> = -{i \over \hbar} <\Psi(q,\t)]|[Q,\A]|\Psi(q,\t)>
\label{q-bracket}
\eeq
The correspondence between \rf{q-bracket} and
\rf{c-bracket} in the semiclassical limit can be studied using
the WKB approximation.  In this approximation, to leading order
in $\hbar$, a solution of $\A \Phi = -\E \Phi$ and all other
constraints \rf{Econstraints} has the form
\beq
      \Phi_{\E,q'}(q) = \exp\left[i\sqrt{\E} S_{HP}[q,q']/\hbar \right]
\label{WKB}
\eeq
where
\beq
        S_{HP} \equiv {S^\E_{HP} \over \sqrt{\E}}
\eeq
is Hamilton's principal function for $\E=1$.  Of course, eq. \rf{WKB}
is only valid away from caustics, in the classically allowed region.
We can now write a general $\t$-dependent WKB solution as a
wavepacket of the form
\bea
     \Psi(q,\t) &=& \int d\E Dq' \; F[\E,q'] \exp[i\tilde{S}]
\non \\
            &=&  \int d\E Dq' \; F[\E,q']
\exp \left[i\{ \E \t + \sqrt{\E} S_{HP}[q,q'] + \theta[q']\}/\hbar \right]
\label{WKBpacket}
\eea
where $F[\E,q']$ and $\theta(q')$ are real-valued.  To make contact with
classical physics, suppose that $F$ is sharply peaked
around some value $\E_0$ and configuration $q^a_0(x)$, and define
\beq
        p_{0a}(x) \equiv \left({\d \theta[q'] \over \d q'^a(x)}
\right)_{q'=q_0}
\eeq
Then, according
to the principle of constructive interference \cite{Gerlach},
the wavefunction $\Psi(q,\t)$ will be peaked around configurations
$q^a(x)$ such that the phase $\tilde{S}$ in \rf{WKBpacket}
is stationary w.r.t. small variations of the parameters
$\E,q'$ around $\E_0,q^a_0(x)$.  In other words,
$q^a(x)$ satifies
\bea
\left({\partial \tilde{S} \over \partial \E}\right)_{\E=\E_0} &=& 0
{}~~~~\Rightarrow~~~~ \t = -{1 \over 2 \sqrt{\E_0} } S_{HP}[q,q']
\non \\
\left({\d \tilde{S} \over \d q'^a}\right)_{q'=q_0} &=& 0
{}~~~~\Rightarrow~~~~ p_{0a} = -\sqrt{\E_0} \left(
{\d S_{HP} \over \d q'^a}\right)_{q'=q_0}
\label{statphase}
\eea

   The principle of constructive interference is the standard way
to make the connection between WKB wavefunctions and
the classical equations of motion \cite{Gerlach}, \cite{MTW}.
Equations \rf{statphase}, with $S_{HP}$ satisfying the Hamiltonian-Jacobi
equations \rf{HJeq} and \rf{HJcon},
are sufficient to specify a classical solution
$q^a(x,\t)$ beginning from an initial configuration $q^a(x,0)=q_0^a(x)$,
and initial momenta $p_{0a}(x)$ satisfying the appropriate constraint
equations \rf{HJcon}.

   The first of the equations in \rf{statphase} tells us that,
for WKB wavefunctions of the form \rf{WKBpacket}, the "wavefunction
of the Universe" is not peaked around a single 3-manifold at time
$\t$, but is instead peaked at configurations (3-manifold + fields)
$q^a(x,\t)$ in superspace such that the classical action between the initial
configuration $q_0$ and $q^a$ is a constant, i.e.
\bea
          \t &=& -{1 \over 2 \sqrt{\E_0}} S_{HP}[q,q_0]
\non \\
             &=& -{1 \over 2 \E_0} S^{\E_0}_{HP}[q,q_0]
\label{time}
\eea
Therefore, even in the classical limit, $\t$ does not in general single
out a {\it particular} 3-manifold and set of fields, obeying \rf{time}.
The second equation in \rf{statphase} is equivalent to the dynamical
equations of the classical theory, and ensures that the wavefunction
is peaked only at configurations $q^a$ obtained from spacelike
slices of the classical solution determined by an initial
$\{q_0^a,p_{0a}\}$.

    The situation for $F[\E,q']$ concentrated at some $\E_0$ and $q_0$
is summarized in Fig. 2.  At a given value of the evolution parameter
$\t$, the wavefunction of the Universe is peaked at a whole class of
configurations, which correspond to spacelike slices of the classical,
4-dimensional solution, satisfying the restriction \rf{time}.

   Now after a measurement of some observable, it is expected that
the Universe should be left in an eigenstate, or
approximate eigenstate, of that
observable immediately after the measurement.  This is a great difficulty
for the standard formulation of canonical quantum gravity, since an
eigenstate of, e.g., 3-geometry is not a solution of the
Wheeler-de Witt equation,
and therefore not a physical state.  In our formulation, the situation
is more favorable.  Imagine that some measurement were performed which
determined the values of the observables $\{q^a,p_a\}$, or some subset
of these observables (modulo three-dimensional diffeomorphisms)
up to the limits imposed by the $\D q \D p$ uncertainty principle.
Immediately after the measurement, the Universe would be left in a
state of the form \rf{WKBpacket} with $\t=0$, and $q_0=q_{observed}$
(modulo three-dimensional diffeomorphisms).  This means that,
at $\t=0$, the wavefunction of the Universe would be peaked around
configurations $q^a$ satisfying
\beq
         S_{HP}[q,q_{observed}] = 0
\label{qq}
\eeq
Since: (i) the action $S_{HP}[q,q']$ of a classical solution bounded by
$q$ and $q'$ is only well-defined if $q'$ is a Cauchy surface for
$q$ (i.e. if $q$ and $q'$ do not intersect on the 4-dimensional
solution manifold); and noting that (ii) the classical
action is monotonic in the evolution parameter (see eq. \rf{DSdt}),
away from caustics/turning points;
eq. \rf{qq} implies that the wavefunction at $\t=0$
is peaked only at configurations which are equivalent, up to
three-dimensional diffeomorphisms, to
\beq
         q^a(x) = q^a_{observed}(x)
\eeq
In other words, there exist physical states which are at least
approximate eigenstates of quantities such as 3-geometry, and
this is made possible by a dispersion in the value of $\E$.

\paragraph{~~$\bullet$ ~~ \bold The "Energy-Time"
Uncertainty Relation}

\indent

\bigskip

      As discussed at length in section 2, the quantity $\E$ is classically
irrelevant: it does not appear in the Euler-Lagrange equations, and therefore
cannot be determined from a classical orbit in superspace.  In fact, it is easy
to see that an uncertainty in $\E$ can be interpreted as an uncertainty in
the effective value of Planck's constant.  This is because in the constraint
equation \rf{WD2} $\E$ can simply be absorbed into a redefinition of
$\hbar$, i.e.
\beq
\left\{ -\hbar_{eff}^2  \k^2 G^{ab} {\d^2 \over \d q^a \d q^b}
+ \rg U \right\}
\Phi_{\E} = 0
\eeq
where
\beq
        \hbar_{eff} = {\hbar \over \sqrt{\E}}
\label{heff}
\eeq
Whatever the correct operator-ordering may be, the absorbtion of $\E$ into
$\hbar$ is always possible.

      We would like to have a quantitative estimate relating
the uncertainty in $\E$ (or $\hbar_{eff}$)
to the spread of wavefunctions along a classical
trajectory (in superspace) around $q^a_{observed}(x)$, leading
to an analog of the time-energy uncertainty relation in ordinary
quantum mechanics.  The discussion below may  serve
to illustrate some of the issues and ambiguities involved in that estimate.

    Consider a WKB wavepacket with $F[\E,q']$ of the form
\beq
     F[\E,q'] = {1 \over 2\r} f[q']
\exp\left[-{(\r - \r_0)^2 \over (\D \r)^2}\right]
\eeq
where  $\r \equiv \sqrt{\E}$.  Then, integrating over $\E$
in \rf{WKBpacket} we have, at $\t=0$,
\beq
     \Psi(q,0) = \int Dq' f[q'] e^{i(\th[q']+\r_0 S_{HP}[q,q'])/\hbar}
\exp[-(\D \r)^2 S_{HP}^2[q,q']/4\hbar^2]
\eeq
Assuming, as before, that $f[q']$ is peaked (modulo diffeomorphisms)
at some configuration $q^a_0(x)$, the integrand has a stationary
phase along the classical manifold statisfying
the second of equations \rf{statphase}.
Consider a configuration $q_{cl}$ along the classical manifold.  Away
from $q_{cl}=q_0$, there will be a suppression factor in the wavefunction
\bea
     |\Psi(q_{cl},0)|^2 &\sim& \exp[-(\D \r)^2 S_{HP}^2[q_{cl},q_0]/2\hbar^2]
\non \\
               &\sim& \exp\left[-\left({\D \E \over \E_0}
S^{\E_0}_{HP}(q_{cl},q_0) \right)^2 / 8\hbar^2 \right]
\eea
If we ask only for the spread of the wavefunction along the classical
manifold, then clearly
\beq
  {\D \E \over \E_0} S^{\E_0}_{HP}(q_{cl},q_0) \sim \hbar
\eeq
or, taking into account eq. \rf{DSdt},
\beq
        \D \E \D \t \sim \hbar
\eeq
which has the form of the usual energy-time uncertainty relation.
However, since $\t$ is just an evolution parameter rather than,
e.g., a proper time interval, this relation is not very informative.

   Let us consider the probability density $P(\D S)$ for being at
{\it any} configuration $q_{cl}$ such that
\beq
       \D S = S_{HP}[q_{cl},q_0]
\eeq
In that case, the measure $\m(\D S)$ of such configurations along the
classical solution manifold becomes important:
\beq
        P(\D S) \sim \m(\D S)\exp\left[-\left({\D \E \over \sqrt{\E_0}}
\D S \right)^2 / 8\hbar^2 \right]
\eeq
In the absence of a regulator for general relativity, we don't really
know the measure $\m(\D S)$.  But let us imagine that it has some
simple power-law dependence on the number of degrees of freedom
$N_p$
\beq
      \m(\D S) \sim (\D S)^{\a N_p}
\label{mDS}
\eeq
where $\a$ is a constant of $O(1)$.  In that case, $P(\D S)$
would be a maximum at
\beq
     \left({\D \E \over \E_0} \D S \right)^2 = 4 \a \hbar_{eff}^2 N_p
\eeq
To relate $\D S$ to an increment of proper time $\D s$ along the classical
manifold, let us consider a matter-dominated Friedman universe,
in which
\beq
       \D S \sim M \D s
\eeq
where $M$ is the total mass, and $\D s$ is the average increment of
proper time between $q_0$ and $q_{cl}$ in a synchronous coordinate
system.  Then we have
\beq
 \left({\D \E \over \E_0} \D s \right)^2 \sim  \a \hbar_{eff}^2
{N_p \over M^2}
\eeq
Let $V$ be the volume and $\r_m$ the density of matter in the Universe,
and denote by $v_0$ the volume per degree of freedom. Then
\beq
 {\D \E \over \E_0} \D s \sim  {\hbar_{eff} \over \r_m} \sqrt{\a \over v_0 V}
\eeq
or, equivalently,
\beq
 {\D \hbar_{eff} \over \hbar_{eff}} \D s \sim
 \hbar_{eff} \sqrt{\a \over \r_m^2 v_0 V}
\label{et}
\eeq
which is the analog, in our formulation, of the energy-time uncertainty
relation of non-parametrized theories.

   It is difficult to assign a reliable number to the right hand side
of expression \rf{et}, even assuming the validity of eq. \rf{mDS} as a
measure of configurations, which was used in the derivation.
The mass density $\r_m$ can be taken as, roughly, the critical density
for closing the Universe ($\sim 10^{-26}~\mbox{kg/m}^3$), and
$v_0$, assuming it is different from zero, is presumably the Planck volume
($\sim 10^{-105}~\mbox{m}^3$).  But there is
still tremendous uncertainty in the volume $V$ of the Universe.
One could even speculate that, in an appropriate continuum limit of
regularized quantum gravity, $v_0 \ra 0$ and $V \ra \infty$ such that
$v_0 V$ tends to a finite constant.  In any case, while it appears that
the transfer matrix formulation must generate some uncertainty relation
between proper time and the fractional uncertainty in Planck's
constant, we are so far unable to say anything quantitative.

\section{Non-Stationary States in Minisuperspace}

   We will now illustrate the formalism developed above in the
simplest context possible: the quantum mechanical time-evolution
of a homogeneous, isotropic Friedmann universe filled with a relativistic
perfect fluid.  In this toy model it is possible to "do everything";
i.e. to find the integration measure and operator ordering, to
solve for the spectrum and exact eigenstates of the $\A$ evolution operator,
to form wavepackets with a conserved norm, and to track the evolution of
such "wavefunctions of the Universe" through collapse towards the
singularity, "bounce", expansion, and recollapse.

  The Friedmann universe is described by the metric
\beq
ds^2=\s^2[-N^2dt^2+a^2d\Omega_3^2]
\label{4metric}
\eeq
where $N$ and $a$ are the lapse and scale factors and
$\s^2={2G_{N}\over 3\pi}$.
The perfect fluid content of this homogeneous and isotropic universe is
specified by the energy-momentum tensor
\beq
T_{\m\n}=(\r+p)U_{\m}U_{\n}+p g_{\m\n}
\label{tensor}
\eeq
where $g_{\m\n}$ is the Friedmann metric corresponding to the line
element \rf{4metric}, $U_{\m}=(1,0,0,0)$ with $U_{\m}U^{\m}=-1$
is the four velocity of the fluid and $\r$ and $p$ are, respectively, the
energy density and pressure of the fluid.

For a perfect fluid described by the equation of state
\beq
p=(\g-1)\r
\label{ellis}
\eeq
the action for gravity and matter (with zero bare cosmological constant)
which is consistent with the definition
of the energy tensor \rf{tensor} (see, e.g., \cite{Ryan}) can be written
as
\beq
S_{bm}={1\over 2}\int dt\left [-{a\dot a^2\over N}+Na-
Na_0^{3\g-2}a^{3(1-\g)}\right ]
\label{4action}
\eeq
Classically, these models represent, for $\g>2/3$, a universe which expands
up to a maximum radius $a_0$ and then recollapses towards the singularity
at $a=0$ and,
for $\g<2/3$, an inflationary universe expanding from the minimum radius
$a_0$ (see, e.g., \cite{Ellis}).\footnote{Also note that we must have
$\g\leq 2$ for the
sound wave velocity of the fluid to be less than the speed of light.}

The perfect fluid does not introduce any extra
dynamical field in this model, and it is easy to rewrite
the action \rf{4action} in the Hamiltonian form corresponding to
the time-parametrized theory of eq. \rf{St}.
The only degrees of freedom are given by the scale factor ($q^a=a$) and its
conjugate momentum $p_a=-{a\dot a\over N}$, and the Hamiltonian $\H$
is given by eq. \rf{St} with
\bea
G_{aa}&=&-a
\non \\
V(a)&=&{a\over 2}\left [\left( {a\over a_0}\right)^{2-3\g}-1\right ]
\label{bulk2}
\eea
(and $m_0=1$).
The `superpotential' $V$ is positive definite in the classically allowed
region $a<a_0$ for $\g>2/3$ ($a>a_0$ for $\g<2/3$), and it changes
sign at the classical turning point $a=a_0$.

Following the discussion of section 3, we can immediately write the
expression for the evolution operator $\A$ in the 1-d minisuperspace
(see eq. \rf{Hmini}) as
\beq
\A =-{\hbar^2\over \vert\G\vert^{1/2}}{\pa\over \pa a}\vert\G\vert^{1/2}
\G^{aa}{\pa \over \pa a}
\label{bulk1}
\eeq
In this case there is no operator ordering term involving $\cal{R}$
since, obviously,
a one dimensional superspace has vanishing scalar curvature ${\cal{R}}=0$.
More explicitly, using  $\G_{aa}=\G=-2aV(a)$ and eq. \rf{bulk2},
we can write
\bea
\A &=&\hbar^2
[2a\vert V\vert]^{-1/2}{\pa \over \pa a}{[2a\vert V\vert]^{1/2}\over
2a V}{\pa \over \pa a}=
\non \\
&=&
{\hbar^2\over 2a V}\left [{\pa^2\over \pa a^2}-{1\over 2a}\left (1+{a\over V}
{\pa V\over \pa a}\right ){\pa \over \pa a}+\d (a-a_0)
\mbox{sign}\left(V{\pa V\over
\pa a}\right ){\pa \over \pa a}\right ]
\label{bulk3}
\eea
The problem is now to solve for all the stationary states
\beq
\A\Phi_{\E}[a]=-\E\Phi_{\E}[a]
\label{bulk}
\eeq
each of which is a solution of a Wheeler-de Witt equation with
a particular operator-ordering, and an effective value of Planck's constant
which depends on $\E$ (see eq. \rf{heff}).  With these solutions in hand,
we can then construct non-stationary states as a linear superposition
\beq
\Psi[a, \t]= \sum_{\E}a_{\E}\Phi_{\E}[a]e^{i\E \t/\hbar}
\eeq

   Since the evolution operator $\A$ of \rf{bulk3} is very singular
at the classical turning point $a=a_0$ (the second line of eq. \rf{bulk3}
does not even make sense, since the product of two distributions is
not defined) we shall consider the regions where $V>0$ and $V<0$
separately, and then impose some appropriate junction condition at the
classical turning point $a=a_0$.

   Let us first consider the region $V>0$.
In this region we change from the coordinate $a$ to the `tortoise' coordinate
\beq
a_+ \dot =\pm\int_{a_0}^a d\bar a~[2V \bar a]^{1/2}=\pm\int_{a_0}^a d\bar
a ~\bar a
\left [\left ({\bar a\over a_0}\right )^{2-3\g}-1\right ]^{1/2}
\label{bulk4}
\eeq
where the plus sign is for the case $\g<2/3$ and the minus sign for $\g>2/3$.
The coordinate $a_+$ is chosen such as to start from zero at $a=a_0$ and
to be semi-positive definite, monotonically growing to infinity as $a>a_0
\rightarrow \infty$ ($\g<2/3$) or to the maximum $a_{+ M}>0$ as
$a\rightarrow 0$ ($\g>2/3$):
\bea
a_+&\in & [0, \infty)~~~~,~~~~a>a_0~~~~,~~~~\g<2/3
\non \\
a_+&\in & [0, a_{+ M}]~~~~,~~~~a<a_0~~~~,~~~~\g>2/3
\non \\
a_+&\geq&0~~~~,~~~~a_+(a_0)=0~~~~,~~~~a_+(0)=a_{+ M}
\label{bulk5}
\eea
With this choice, the eigenvalue problem for the evolution operator simply
becomes
\beq
\hbar^2{\pa^2\over \pa a_+^2}\Phi^+_{\E}[a_+]=-\E\Phi^+_{\E}[a_+]
\label{bulk6}
\eeq
which is the Schr\"odinger problem for the motion of a free particle
with energy $\E$.
Assuming from now on that $\E>0$,\footnote{The choice $\E>0$ is
required in order to have a real-valued Planck's constant.}
the general solution of eq.
\rf{bulk6} in the region $V>0$ is clearly a combination of plane waves
\beq
\Phi^+_{\E}[a_+]=A~e^{i\sqrt{\E}a_+/\hbar}+B~e^{-i\sqrt{\E}a_+/\hbar}
\label{bulk7}
\eeq
Similarly, in the region $V<0$, we transform to the `tortoise' coordinate
\beq
a_-\dot =\pm\int_{a_0}^a d\bar a~[-2V \bar a]^{1/2}=\pm\int_{a_0}^a
d\bar a ~\bar a
\left [1-\left ({\bar a\over a_0}\right )^{2-3\g}\right ]^{1/2}
\label{bulk8}
\eeq
where, again, the plus sign is for the case $\g<2/3$ and the minus sign
for $\g>2/3$.
The coordinate $a_-$ is zero at $a=a_0$ and negative semi-definite,
monotonically growing to minus infinity as $a>a_0
\rightarrow \infty$ ($\g>2/3$) or to the minimum $a_{- m}<0$ as
$a\rightarrow 0$ ($\g<2/3$):
\bea
a_-&\in & [a_{- m}, 0]~~~~,~~~~a<a_0~~~~,~~~~\g<2/3
\non \\
a_-&\in & (-\infty, 0]~~~~,~~~~a>a_0~~~~,~~~~\g>2/3
\non \\
a_-&\leq&0~~~~,~~~~a_-(a_0)=0~~~~,~~~~a_-(0)=a_{- m}
\label{bulk9}
\eea
In this case, eq. \rf{bulk} becomes
\beq
\hbar^2{\pa^2\over \pa a_-^2}\Phi^-_{\E}[a_-]=\E\Phi^-_{\E}[a_-]
\label{bulk10}
\eeq
with solutions in the (classically forbidden) $a_-$ region
\beq
\Phi^-_{\E}[a_-]=C~e^{\sqrt{\E}a_-/\hbar}+D~e^{-\sqrt{\E}a_-/\hbar}
\label{bulk11}
\eeq

Let us now build the general solution of the original eigenvalue problem
\rf{bulk} in the whole range $a\in [0, \infty)$.

First, we must impose a junction condition at the turning point
$a_+=a_-=0$ ($a=a_0$).
We multiply both sides of eq. \rf{bulk} by $[2a\vert V\vert]^{1/2}$ and,
using eq. \rf{bulk3}, we
integrate around the turning point $a_0$.
Assuming a reasonably smooth
behaviour of $\Phi_{\E}$ at $a_0$, we find that
\beq
\lim_{\d\ra 0}\int_{a_0-\d}^{a_0+\d}da~{\pa \over \pa a}{[2a\vert V\vert]^{1/2}
\over 2aV}{\pa \over \pa a}\Phi_{\E}=-\lim_{\d\ra 0}\int_{a_0-\d}^{a_0+\d}
da~[2a\vert V\vert]^{1/2}{\E\over \hbar^2}\Phi_{\E}=0
\label{bulk12}
\eeq
Evaluating the integral on the left hand side, the joining condition on
first derivatives of $\Phi_{\E}$ in the $a$ coordinate can be rewritten as
\beq
{\vert V\vert^{1/2}\over V}{\pa \Phi_{\E}\over \pa a}\biggr \vert_{a_{0^+}}=
{\vert V\vert^{1/2}\over V}{\pa \Phi_{\E}\over \pa a}\biggr \vert_{a_{0^-}}
\label{bulk13}
\eeq
Both $\Phi_{\E}$ and its derivative must be continuous at $a=a_0$.

When we turn to $a_\pm$ coordinates, it is easy to see that
first derivatives must be `discontinuous', i.e. we have the junction conditions
\bea
\Phi^+_{\E}[0]&=&\Phi^-_{\E}[0]
\non \\
{\pa \Phi^{\pm}_{\E}\over \pa a_{\pm}}\biggr \vert_{a_{0^+}}&=&
-{\pa \Phi^{\mp}_{\E}\over \pa a_{\mp}}\biggr \vert_{a_{0^-}}
\label{bulk15}
\eea
where in the last equation upper signs refer to the case $\g<2/3$,
and lower signs to the case $\g>2/3$.

Second, the evolution operator is defined on the `half line' $a>0$, and
therefore it is not essentially self-adjoint in $L^2[0, \infty)$ with the
measure $\sqrt{\vert \G\vert}$.
One can build, however, a one-parameter family of
self-adjoint extensions (which
guarantee norm conservation and unitarity) by
appropriately choosing boundary conditions at the origin (see, e.g.,
\cite{Reed}).
In the $a_\pm$ coordinates this is translated into the condition
\beq
{\pa \Phi^{\pm}_{\E}\over \pa a_{\pm}}\biggr \vert_{a_{\pm, m(M)}}=
\a\Phi^{\pm}_{\E}\vert_{a_{\pm, m(M)}}
\label{bulk16}
\eeq
where $\a$ is an arbitrary real parameter with range $(-\infty, \infty)$.
The lower signs (and index $m$) correspond to the case $\g<2/3$, while
the upper signs (and index $M$) correspond to $\g>2/3$.

Self-adjointness of the evolution operator $\A$ automatically guarantees
orthogonality of eigenfunctions, i.e.
\bea
\int_0^{\infty}da~\sqrt{\vert\G\vert}\Phi^{\ast}_{\E_1}[a]\Phi_{\E_2}[a]
&=&\int_{a_{- m}(-\infty)}^0da_-~{\Phi^-_{\E_1}}^{\ast}[a_-]\Phi^-_{\E_2}
[a_-]~
\non \\
&+&\int_0^{\infty(a_{+ M})}da_+~{\Phi^+_{\E_1}}^{\ast}[a_+]\Phi^+_{\E_2}[a_+]
=\d(\E_1, \E_2)
\label{bulk30}
\eea
where $\d(\E_1, \E_2)$ is the Dirac delta function when the spectrum is
continuous and the Kronecker delta for a discrete spectrum (and
integration limits in parenthesis are for the case $\g>2/3$).

It is also possible to show that the system of eigenfunctions is complete,
i.e., defining $\rho\dot =\sqrt{\E}$, that
\beq
\sum_n\Phi_{\rho_n}^{\ast}[a]\Phi_{\rho_n}[a^{\prime}]
={\d(a-a^{\prime})\over
\sqrt{\vert\G\vert}}
\label{bulk31}
\eeq
(where of course the sum is meant as an integral for a continuous
spectrum).

   We are now in the position to write the  exact eigenfunctions for
different $\g$ as a function of the scale factor $a$.

\paragraph{~~$\bullet$ ~~ \bold ${\g<2/3}$}
\indent

In the case $\g<2/3$, the spectrum is continuous and non degenerate, and
from conditions \rf{bulk15}-\rf{bulk16}
one finds, for a generic boundary condition at the origin $a=0$,
\bea
A&=&\left [(1+i) +(1-i){(\sqrt{\E}-\hbar\a)
\over (\sqrt{\E}+\hbar\a)}e^{2\sqrt{\E}
a_{- m}/\hbar}\right ]{C\over 2}
\non \\
B&=&\left [(1-i) +(1+i){(\sqrt{\E}-\hbar\a)
\over (\sqrt{\E}+\hbar\a)}e^{2\sqrt{\E}
a_{- m}/\hbar}\right ]{C\over 2}
\non \\
D&=&{(\sqrt{\E}-\hbar\a)\over (\sqrt{\E}+\hbar\a)}e^{2\sqrt{\E}a_{-
m}/\hbar}C
\label{bulk17}
\eea

For instance, in the case $\g =0$, which classically corresponds to a
de Sitter universe expanding from the minimum radius $a_0$,
the `tortoise' coordinates derived from eqs. \rf{bulk4} and \rf{bulk8} are
\bea
a_-&=&-{a_0^2\over 3}\left [1-\left ( {a \over a_0}\right )^2\right ]^{3/2}
\non \\
a_+&=&{a_0^2\over 3}\left [\left ( {a \over a_0}\right )^2-1\right ]^{3/2}
\non \\
a_{- m}&=&-{a_0^2\over 3}
\label{bulk18}
\eea
Moreover, if we choose boundary conditions at the origin such that
$\Phi_{\E}(0)=0$ (or, equivalently, from eq. \rf{bulk16}, such that
$\a=\infty$), the eigenfunctions of $\A$ are
\beq
\Phi_{\E}[a]=
C\left\{ e^{-{\sqrt{\E}a_0^2\over 3\hbar}\left[1-\left ({a\over a_0}\right )^2
\right]^{3/2}}-e^{{\sqrt{\E}a_0^2\over 3\hbar}\left[\left[1-\left ({a\over a_0}
\right )^2\right]^{3/2}-2\right]}\right \}
\label{bulk19}
\eeq
for $a<a_0$ and
\bea
\Phi_{\E}[a]&=&
2Ce^{-{\sqrt{\E}a_0^2\over 3\hbar}}\Biggl\{
\sinh\left[{\sqrt{\E}a_0^2\over 3\hbar}\right]
\cos\left [{\sqrt{\E}a_0^2\over 3\hbar}\left[\left ({a\over a_0}\right )^2-1
\right]^{3/2}\right]
\non \\
&-&\cosh\left[{\sqrt{\E}a_0^2\over 3\hbar}\right]
\sin\left [{\sqrt{\E}a_0^2\over 3\hbar}\left[\left ({a\over a_0}\right )^2-1
\right]^{3/2}\right]\Biggr\}
\label{bulk20}
\eea
for $a>a_0$.
With this choice of boundary conditions, the eigenfunction is real for all
values of $a$.

\paragraph{~~$\bullet$ ~~ \bold ${\g>2/3}$}
\indent
\bigskip

  The case $\g>2/3$ corresponds classically to
a universe which expands from the singularity at $a=0$ to the
maximum radius at $a=a_0$, and then collapses back towards the
singularity.
Since now the range of the `tortoise' coordinate $a_-$ is infinite (see
eq. \rf{bulk9}), to ensure square integrability of the eigenfunctions
(eq. \rf{bulk30}) clearly we must have $D=0$.
Therefore, due to conditions \rf{bulk15} and \rf{bulk16},
the spectrum is discrete and non degenerate.

Normalizing the eigenfunctions to one in the range $a\in [0, \infty)$,
from eqs. \rf{bulk15} and \rf{bulk30} we have that
\bea
A&=&{(1+i)\over 2}C
\non \\
B&=&{(1-i)\over 2}C
\non \\
C&=&\left[{2\sqrt{\E}\over \hbar\cos(2\sqrt{\E}a_{+ M}/\hbar)
+2\sqrt{\E}a_{+ M}}
\right ]^{1/2}e^{i\theta}
\label{bulk21}
\eea
where $\theta$ is an arbitrary phase.

Moreover, imposing a generic boundary condition at the origin according to
eq. \rf{bulk16} gives an implicit expression for the discrete spectrum
of $\E$, i.e.
\beq
\tan(\sqrt{\E}a_{+ M}/\hbar)={\hbar\a+\sqrt{\E}\over\hbar\a-\sqrt{\E}}
\label{bulk23}
\eeq
For example, in the case of a pressureless, dust-dominated universe
with $\g=1$, the explicit form of the `tortoise' coordinates is
\bea
a_-&=&{a_0^2\over 4}\left [\mbox{arccosh}\left({a\over a_0}\right )^{1/2}+
\left (1-{2a\over a_0}\right)\left [{a \over a_0}\left({a\over a_0}-1\right
)\right]^{1/2}\right ]
\non \\
a_+&=&{a_0^2\over 4}\left [\mbox{arccos}\left({a\over a_0}\right )^{1/2}+
\left (1-{2a\over a_0}\right)\left [{a \over a_0}\left(1-{a\over a_0}\right
)\right]^{1/2}\right ]
\non \\
a_{+ M}&=&{\pi\over 8}a_0^2
\label{bulk22}
\eea
Choosing boundary conditions such that $\Phi_{\E}(0)=0$ (or $\a =\infty$),
from eq. \rf{bulk23} one can explicitly write the discrete eigenvalues for the
dust-dominated universe as
\beq
\sqrt{\E}={2(4n+1)\hbar\over a_0^2}~~~~,~~~~n=0, 1,...
\label{bulk24}
\eeq
and the $\A$ eigenfunctions are
\bea
\Phi_{\E}[a]&=&{2^{3\over 2}e^{i\theta}\over \pi^{1\over 2}a_0}
\Biggl\{\cos\left [
{\sqrt{\E}a_0^2\over 4\hbar}\left[\mbox{arccos} \left({a\over a_0}\right)^{1
\over 2}+
\left(1-{2a\over a_0}\right)\left[{a\over a_0}\left(1-{a\over a_0}\right)
\right]^{1\over 2}\right]\right]
\non \\
&-&\sin\left[{\sqrt{\E}a_0^2\over 4\hbar}\left[\mbox{arccos}
\left({a\over a_0}\right)
^{1\over 2}+\left(1-{2a\over a_0}\right)\left[
{a\over a_0}\left(1-{a\over a_0}\right)\right]^{1\over 2}
\right]\right]\Biggr\}
\label{bulk25}
\eea
for $a<a_0$ and
\beq
\Phi_{\E}[a]={2^{3\over 2}e^{i\theta}\over \pi^{1\over 2}a_0}\exp
\left\{{\sqrt{\E}a_0^2\over 4\hbar}
\left[\mbox{arccosh} \left({a\over a_0}\right)
^{1\over 2}+\left(1-{2a\over a_0}\right)\left[
{a\over a_0}\left({a\over a_0}-1\right)\right]^{1\over 2}
\right]\right\}
\label{bulk26}
\eeq
for $a>a_0$.

  In the same way, one can solve for the stationary states of a
radiation-dominated universe with $\g=4/3$.
In this case the discrete eigenvalues are
\beq
\sqrt{\E}={(4n+1)\hbar\over a_0^2}
\label{bulk27}
\eeq
and the $\A$-eigenfunctions are
\bea
\Phi_{\E}[a]&=&{2e^{i\theta}\over \sqrt{\pi}a_0}\Biggl\{\cos\left [
{\sqrt{\E}a_0^2\over 2\hbar}\left[\mbox{arccos} \left({a\over a_0}\right)-
{a\over a_0}\left[1-\left({a\over a_0}\right)^2\right]^{1/2}
\right]\right]
\non \\
&-&\sin\left[{\sqrt{\E}a_0^2\over 2\hbar}\left[\mbox{arccos}
\left({a\over a_0}\right)
-{a\over a_0}\left[1-\left ({a\over a_0}\right)^2\right]^{1/2}
\right]\right]\Biggr\}
\label{bulk28}
\eea
for $a<a_0$ and
\beq
\Phi_{\E}[a]={2e^{i\theta}\over \sqrt{\pi}a_0}\exp
\left[{\sqrt{\E}a_0^2\over 2\hbar}\left[\mbox{arccosh}
\left({a\over a_0}\right)
-{a\over a_0}\left[\left({a\over a_0}\right)^2-1\right]^{1/2}
\right]\right]
\label{bulk29}
\eeq
for $a>a_0$.
The main properties of the wave function are essentially the same as those
of the dust-dominated universe.

    Note that in both the $\g>2/3$ and $\g<2/3$ cases,
the $\A$ eigenstates are real-valued.  In the more conventional
"Born-Oppenheimer" interpretation \cite{Banks} of the
Wheeler-de Witt equation, such wavefunctions represent a superposition of
`collapsing' and `expanding' universes in the classically allowed
region.
In our formulation, by contrast, such states are stationary, and dynamics
(such as the expansion and collapse of the universe) arises from their
superposition.  We will now illustrate such dynamics by constructing
a non-stationary state from the $\A$ eigenstates.

Let us consider the case of a $\g>2/3$ Friedman universe, which has
a discrete spectrum for $\E$, and oscillating eigenfunctions
in the classically allowed range
$a\in [0, a_0]$.
The general form of the wave packet for an arbitrary distribution
$f[\rho_n]$
of "momenta" $\rho_n\dot =\sqrt{\E_n}$ is given by the standard formula
\beq
\Psi[a, \t]=\sm f[\rho_n]\Phi_{\rho_n}[a]e^{i\rho_n^2\t/\hbar}
\label{bulk32}
\eeq
where
\beq
f[\rho_n]=\int_0^{\infty}da~\sqrt{\vert\G\vert}\Psi[a, 0]
\Phi^{\ast}_{\rho_n}[a]
\label{bulk33}
\eeq
and, for the choice of boundary conditions $\Phi_{\E}(0)=0$,
\beq
\rho_n={\pi(4n+1)\hbar\over 4a_{+ M}}
\label{bulk34}
\eeq
We build up the wave packet such that it is initially localized in the
`classically allowed' region $a\in [0, a_0]$ and it moves towards the
singularity with "momentum" $\rho_n=\sqrt{\E_n}$ centered around $\rho_{n_0}$.
For our example, we choose an initial wavepacket which is a triangle
of unit height, with base of length $2\d$ centered at $a_{+ 0}$, i.e.
\beq
\Psi[a_+, 0]=e^{i\rho_{n_0}a_+/\hbar
}\left [1-{\vert a_+-a_{+ 0}\vert\over \d}\right]
[\theta(a_+-a_{+ 0}+\d)-\theta(a_+-a_{+ 0}-\d)]
\label{bulk35}
\eeq
where $\theta$ is the Heaviside step function.

Since the initial packet has support only in the region $a_+$,
from eqs. \rf{bulk7}, \rf{bulk33} and \rf{bulk35} we find
\bea
f[\rho_n]&=&\int_0^{a_{+ M}}da_+~\Psi[a_+, 0]\Phi_{\rho_n}^{\ast}[a_+]
\non \\
&=&{2\hbar^2\over \d}\biggl\{{A^{\ast}\over (\rho_{n_0}-\rho_n)^2}e^{i(
\rho_{n_0}-\rho_n)a_{+ 0}/\hbar}[1-\cos[(\rho_{n_0}-\rho_n)\d/\hbar]
\non \\
&+&{B^{\ast}\over (\rho_{n_0}+\rho_n)^2}e^{i(\rho_{n_0}+\rho_n)a_{+
0}/\hbar}[1-\cos[(\rho_{n_0}+\rho_n)\d/\hbar]
\biggr\}
\label{bulk36}
\eea
Inserting this result in eq. \rf{bulk32} and taking the modulus square, we
finally find that the probability distribution of the wave packet
evolves, in the classically allowed region $a_+$, as
\bea
\vert\Psi[a_+, \t]\vert^2&=&{4a^2_{+ M}\over \pi^4\d^2}\Biggl\{
\biggl[\sm [\cos (\m_+(n))+\sin (\s_-(n))]{\sin^2{\pi\d(n-n_0)\over 2a_{+ M}}
\over (n-n_0)^2}
\non \\
&+&\sm [\cos (\m_-(n))-\sin (\s_+(n))]
{\sin^2{\pi\d(n+n_0+1/2)\over 2a_{+ M}}\over (n+n_0+1/2)^2}\biggr]^2
\non \\
&+&\biggl[\sm [\sin (\m_+(n))-\cos (\s_-(n))]{\sin^2{\pi\d(n-n_0)\over 2
a_{+ M}}\over (n-n_0)^2}
\non \\
&+&\sm [\sin (\m_-(n))+\cos (\s_+(n))]
{\sin^2{\pi\d(n+n_0+1/2)\over 2a_{+ M}}\over (n+n_0+1/2)^2}\biggr]^2
\Biggr\}
\label{bulk37}
\eea
where, for convenience of notation, we have defined
\beq
\m_{\pm}(n)={\pi \over a_{+ M}}(n+1/4)\left[\pm(a_+-a_{+ 0})+
{\pi \hbar\over a_{+ M}}(n+1/4)\t\right]
\label{bulk38}
\eeq
and
\beq
\s_{\pm}(n)={\pi \over a_{+ M}}(n+1/4)\left[\pm(a_++a_{+ 0})+
{\pi \hbar\over a_{+ M}}(n+1/4)\t\right]
\label{bulk38bis}
\eeq
Similarly, the probability distribution in the classically forbidden
region $a_-$ is easily found to be
\bea
\vert\Psi[a_-, \t]\vert^2&=&{4a^2_{+ M}\over \pi^4\d^2}e^{{\pi a_-
\over 2a_{+ M}}}\Biggl\{
\biggl[\sm [\cos (\lambda_1
(n))+\sin (\lambda_1(n))]{\sin^2{\pi\d(n-n_0)\over 2a_{+ M}}
\over (n-n_0)^2}e^{{\pi a_-\over a_{+ M}}n}
\non \\
&+&\sm [\cos (\lambda_2(n))-
\sin (\lambda_2(n))]{\sin^2{\pi\d(n+n_0+1/2)\over 2a_{+ M}}
\over (n+n_0+1/2)^2}e^{{\pi a_-\over a_{+ M}}n}\biggr]^2
\non \\
&+&\biggl[\sm [\sin (\lambda_1(n))-\cos (\lambda_1
(n))]{\sin^2{\pi\d(n-n_0)\over 2
a_{+ M}}\over (n-n_0)^2}e^{{\pi a_-\over a_{+ M}}n}
\non \\
&+&\sm [\sin (\lambda_2(n))+\cos (\lambda_2(n))]
{\sin^2{\pi\d(n+n_0+1/2)\over 2a_{+ M}}\over (n+n_0+1/2)^2}
e^{{\pi a_-\over a_{+ M}}n}\biggr]^2
\Biggr\}
\label{bulk39}
\eea
where
\beq
\lambda_1={\pi n\over a_{+ M}}\left [{\pi\hbar
\over 2a_{+ M}}(2n+1)\t-a_{+ 0}\right ]
\label{bulk40}
\eeq
and
\beq
\lambda_2={\pi (2n+1)\over 2a_{+ M}}\left [{\pi \hbar
\over a_{+ M}}n\t+a_{+ 0}\right ]
\label{bulk41}
\eeq

    For our illustrations, we have chosen to compute the probability
distribution for a dust-dominated universe
($\g=1$) with $n_0=400$, $a_0=1$ and
$\d=0.02$ (keeping $10^4$ terms in the series).
The results are plotted in fig. 3, where
different times are labeled by $\t_{\ast}\dot =-10^{5}\t$.
For a comparison, we have also plotted in fig. 4 the modulus of the
eigenfunction of $\A$ for the $\E$ eigenvalue corresponding to the average
$\E$ of the packet, namely $\E_0=(3202)^2$.  This stationary state is the
solution of a Wheeler-de Witt equation with a particular choice of
operator-ordering and Planck's constant.

    Fig. 3 shows the evolution of the wave packet; the packet begins
by moving smoothly towards the singularity while gradually spreading.
As it approaches the classical singularity at $a=0$, the packet starts
oscillating, with the frequency and amplitude of oscillations
increasing up to a maximum which is shown in more detail in fig. 5a.
The wave packet is always zero at $a=0$, consistent with
our choice of boundary conditions.
As $\t$ increases, the packet `bounces' off $a=0$ and moves
back towards the classical turning point at $a=a_0$.
At the point of maximum expansion
the wave packet has again an oscillating behaviour
similar to that close to the classical singularity.
Here, however, part of the wave packet extends into the `classically
forbidden' region $a>a_0$; there is a small, exponentially decaying
probability to find the universe in such a region (see fig. 5b).
At $a=a_0$ both the wave packet and its first derivative are continuous.
As the time parameter continues to increase,
the wave packet `bounces' off $a=a_0$
and moves back towards the classical singularity.
The packet rebounds repeatedly between $a=0$ and $a=a_0$
while gradually becoming delocalized.
It is important to stress (see also \cite{DeWitt}) that nothing pathological
happens to the wave packet as it bounces between the classical singularity
at $a=0$ and the turning point at $a=a_0$. In particular,
there is no infinite compression of the wavefunction, and the
transition from `big crunch' to `big bang' is a (relatively) smooth
process, in this toy model, at the quantum level.

  The norm of the wave packet has been numerically checked to be constant
during all the stages of the evolution, as it obviously should.

  Although we have specialized here to the case of a dust dominated universe,
it is quite clear that we do not expect anything radically different
from the analysis of other cases with $\g>2/3$.
The same holds for the cases $\g<2/3$, the only difference being
that there is only a single bounce of the kind shown in fig. 6, with
the wave packet starting from infinite scale factor, `bouncing'
off the minimum radius $a_0$ and finally going back to infinite scale
factor.
Also, of course, the qualitative picture does not depend critically on
the shape of this initial wavepacket.


   As pointed out in section 2, the space of physical
states is spanned by eigenstates of the evolution operator $\A$ with
different $\E$, and in section 4 it was shown that this can be interpreted as
leading to a quantum indeterminacy of Planck's constant according
to formula \rf{heff}.
In our minisuperspace toy model, it
is possible to directly relate the spread of the wavepacket in
the scale factor $a$ to the spread in $\E$, and thereby to the
dispersion in Planck's constant.  Of course, this model is far
too unrealistic to draw any quantitative conclusions regarding the
dispersion of fundamental constants vs. the spread of the
wavepacket in full quantum gravity.

   To proceed, we construct a wave packet from eigenfunctions of $\A$.
Now, however, we want to introduce a more tractable Gaussian
distribution in the momenta, and to simplify the analysis
let us assume
that the $\E$ eigenvalues are so closely spaced that we can replace
sums over the discrete eigenvalues by an integral,\footnote{This can be
done by replacing
$\sm\rightarrow {a_{+ M}\over 2\pi\hbar}\int_{-\infty}^{\infty}d\rho$ for
$\g>2/3$.}
i.e. we write
\bea
\Psi(a, \t)&=&\int_{-\infty}^{\infty}d\rho
f[\rho]\Phi_{\rho}(a)e^{i\rho^2\t/\hbar}
\non \\
f(\rho)&=&e^{-\b^2(\rho-\rho_0)^2}
\label{bulk42}
\eea
where
\beq
\s_{\rho}^2\dot=(2\b^2)^{-1}
\label{sigmap}
\eeq
is the dispersion in the "momentum" $\rho=\sqrt{\E}$
distribution around the peak at $\rho_0$.

Evaluating the integral in the classically allowed region with
$\Phi_{\rho}=e^{i\rho a_+\over \hbar}$ and taking the modulus, we
easily find that
\beq
\vert \Psi[a, \t]\vert^2\simeq \exp\left \{-{1\over 2}{[a_+(a)+2
\sqrt{\E_0}\t]^2\over \hbar^2\left [\b^2+{\t^2\over \hbar^2\b^2}\right ]}
\right \}
\label{bulk43}
\eeq
where $a_+(a)$ is given by formula \rf{bulk4}.

Following standard analysis (see, e.g., \cite{Messiah}), we see that,
at any given $\t$, the probability distribution of the scale factor
$a$ is peaked around the set of semiclassical trajectories for which
\beq
a_+(a)\vert_{cl}=-2\sqrt{\E_0}~\t
\label{bulk44}
\eeq
or, in other words, for which the phase in the integral \rf{bulk42}
is stationary (compare with eqs. \rf{WKBpacket} and \rf{statphase}).

As an example, we can differentiate both sides of eq. \rf{bulk44} with
respect to $\t$ and find
\beq
{da \over d\t}=-{2\sqrt{\E_0}\over a}\left [\left ({a\over a_0}\right )
^{2-3\g}-1\right ]^{-1/2}
\label{bulk45}
\eeq
If we now compare with the known solutions in classical cosmology
for the perfect fluid matter content (see \cite{Ellis}), i.e.
\beq
{da \over dt}=\left[\left ({a\over a_0}\right )^{2-3\g}-1\right ]^{1/2}
\label{bulk46}
\eeq
we can immediately infer the correspondence between the `internal' time
$t$ and the evolution parameter $\t$ at the classical level, given by
\beq
d\t=-{a\over 2 \sqrt{\E_0}}\left [\left ({a\over a_0}\right )^{2-3\g}-1
\right ]dt
\label{bulk47}
\eeq

We now expand $\vert \Psi[a, \t]\vert^2$ in eq. \rf{bulk43}
around the classical trajectories with $a=a(a_+\vert_{cl})+\D a$
at $\t$ fixed,
and find
\beq
\vert \Psi[a, \t]\vert^2\simeq e^{-{(a-a_{cl})^2\over 2\s_a^2}}
\label{bulk0}
\eeq
where
\beq
\s_a^2(\b, \t)=\left [\b^2+{\t^2\over \hbar^2\b^2}\right ]\left [{\pa a_+\over
\pa a}\right]^{-2}\left({2 \hbar G\over 3\pi}\right)^2
\label{bulk49}
\eeq
and we have reinserted dimensions according to
\beq
a\rightarrow \sqrt{2G\over 3\pi}~a
\label{bulk00}
\eeq
Eq. \rf{bulk0} can be interpreted as saying that the probability
to find the universe in a certain configuration with scale factor $a$
at time $\t$ is given by a Gaussian peaked at the classical trajectory
$a_{cl}$ with a dispersion $\s_a^2$.

Finally, as an exercise, we can tentatively give a lower bound on
the dispersion for the
scale factor by minimizing it with respect to $\b$, and we find
\beq
\{Min [\s_a^2]\}_{\b=\vert \t/\hbar\vert^{1/2}}=\vert a_+\vert
\left [{\pa a_+\over \pa a}\right]^{-2}L_{P, eff}^2
\label{bulk50}
\eeq
where we have defined $L_{P, eff}^2={2\hbar_{eff}G\over 3\pi}$ and the
`dressed' Planck's constant $\hbar_{eff}$ is introduced
according to eq. \rf{heff}.
We can study the behaviour of eq. \rf{bulk50} for some simple examples
of matter content in the universe.
In the case of a de Sitter universe with $\g=0$ we can use formula
\rf{bulk18} for $a_+$, define $a=\xi a_0$ ($\xi\geq 1$) and find
\beq
{\{Min [\s_a^2]\}_{\b=\vert \t/\hbar\vert^{1/2}}\over
L_{P, eff}^2}=
{(\xi^2-1)^{1/2}\over 3\xi^2}<{1\over 36}~~~~,~~~~\forall \xi\geq 1
\label{bulk52}
\eeq
Therefore, if the dispersion in the scale factor is such to be minimized with
respect to $\b$, it will always be negligible for the case $\g=0$.
A similar conclusion can be easily drawn by studying the case of a
radiation-dominated universe.
Unfortunately, this simple analysis does not either provide any reliable upper
bound on $\s_a$.
%

For a dust-dominated universe, instead, using eq. \rf{bulk22} with
$a=\xi a_0$ ($\xi\leq 1$), one has
\beq
{\{Min [\s_a^2]\}_{\b=\vert \t/\hbar\vert^{1/2}} \over L_{P, eff}^2}=
{[\mbox{arccos}(\xi)^{1/2}+(1-2\xi)[\xi(1-\xi)]^{1/2}]\over 4\xi(1-\xi)}
\rightarrow \infty~~,~~\xi\rightarrow 0
\label{bulk56}
\eeq
In this toy model, it is therefore possible to have non-negligible
dispersion close to the
classical singularity, while having $\s_a$ negligible at much later times.

\section{Conclusions}

    In this article we have pointed out that there is a constant
(denoted $\E$) in the classical Hamiltonian of certain parametrized
theories, which cannot be determined from the classical trajectories.
It is the arbitrariness of this constant
in classical physics which allows us to extend the space of physical
states in the corresponding quantum theory, and this is the key to
our proposed resolution of the problem of time.  Solutions of the
Wheeler-de Witt equations are stationary states, in our scheme,
but there is an infinite set of such equations, each with its own parameter
$\E$.  Non-stationary states are composed of superpositions of such
states, with different $\E$ parameters.  It was seen that this formalism
can be derived directly from a transfer-matrix quantization of parametrized
theories.

   As noted in section 4, the constant $\E$ in the Wheeler-de Witt
equation (equivalent to $\A \Phi_\E = -\E \Phi_\E$ in our notation)
can always be absorbed into a redefinition of Planck's constant
$\hbar_{eff} = \hbar/\sqrt{\E}$. In the special case of pure gravity,
the constant $\E$ could alternatively be absorbed into a redefinition
of Newton's constant \rf{GNewt} (as was done in ref. \cite{JG}),
but in pure gravity this
is anyway equivalent to rescaling the Planck constant, since
only the combination $\hbar G/\sqrt{\E}$ appears in the Wheeler-de Witt
equation in this case.  In the general case, any non-stationary state
\beq
       \Psi[q,\t] = \sum_{\E} a_\E \Phi_\E[q] e^{i\E \t/\hbar}
\eeq
involves a superposition of Wheeler-de Witt wavefunctions
$\Phi_\E[q]$, each
with a different effective value of Planck's constant, so in this
sense any non-stationary physical state entails an inherent
uncertainty in the value of Planck's constant.  Unfortunately,
we have not been able to place a reliable lower bound on this
uncertainty, for reasons which were discussed in section 4, above.
Nevertheless, the possibility that there is an inherent uncertainty in the
effective value of Planck's constant, which {\it might} be large enough
to be observable, raises some interesting phenomenological questions,
which we hope to address in the near future.

\vspace{33pt}

\noindent {\Large \bf Acknowledgements}{\vspace{11pt}}

   It is a pleasure to acknowledge the hospitality of the Lawrence Berkeley
Laboratory, where most of this work was carried out.
A.C. was supported by an INFN grant at the Lawrence Berkeley Laboratory.
J.G.'s research is supported in part by  the U.S. Dept. of Energy, under
Grant No. DE-FG03-92ER40711.

\vspace{33pt}
\appendix

\section{Appendix}
We briefly comment here on a 2-d minisuperspace model for a universe
filled with a homogeneous scalar field $\p$ with arbitrary potential
 ${\cal{V}}(\p)$ and with a bare cosmological constant $\l$.
The geometrical ansatz is the same as that of the 1-d minisuperspace
model, i.e. the isotropic and homogeneous Friedmann metric given by
eq. \rf{4metric}.
The reparametrization invariant action for such a model is
\beq
S_{sc}={1\over 2}\int dt\left [-{a\dot a^2\over N}+{a^3\dot \p^2\over N}
 +Na-N(\l+{\cal{V}}(\p))a^3\right ]
\label{scalar0}
\eeq
This can be easily written in the Hamiltonian form of eq. \rf{Smini},
with $p_a=-{a\dot a\over N}$, $p_{\p}={a^3\dot \p\over N}$, $m=1$ and
\bea
G_{aa}&=&-a
\non \\
G_{\p\p}&=&a^3
\non \\
 V(a, \p)&=&{a\over 2}\left [({\cal{V}}(\p)+\l)a^2-1\right ]
\label{scalar2}
\eea
The general form of the evolution operator $\A$ is
\beq
\A =-{3\hbar^2\over 2\vert\G\vert^{1/2}}\left [{\pa\over \pa a}\vert\G
\vert^{1/2}\G^{aa}{\pa \over \pa a}+{\pa\over \pa \phi}\vert\G
\vert^{1/2}\G^{\phi\phi}{\pa \over \pa \phi}\right ]+{\hbar^2{\cal{R}}\over 2}
\label{scalar1}
\eeq
 where, as usual, we have introduced the `supermetric' $\G_{m n}=2VG_{m
n}$, with determinant $\G$ and corresponding scalar curvature
\beq
 {\cal{R}}={1\over 2aV}\left [\left (1+{2V\over a}
\right ){1\over
 V^2}+{2\over a^2}{\pa^2\over \pa \p^2}\ln V
\right ]
\label{scalar3}
\eeq
As a consequence of the hermiticity requirements for the measure, similarly
to the 1-d minisuperspace case, the formal expression of the evolution
 operator is not well defined at the `turning points' where $V=0$.
In fact, from eqs. \rf{scalar2}-\rf{scalar3} and using the explicit
form of the supermetric, one gets
\bea
 \A &=&{3\hbar^2\over 4a V}\biggl \{{\pa^2\over \pa a^2}+{1\over a}{\pa \over
 \pa a}-{1\over a^2}{\pa^2\over \pa \p^2}+2\d (V)sign (V)
 \biggl[{\pa V\over \pa a}{\pa \over \pa a}
\non \\
 &-&{1\over a^2}{\pa V\over \pa \p}{\pa \over \pa \p}\biggr ]
 +{1\over 3}\left [\left (1+{2V\over a}\right ){1\over
 V^2}+{2\over a^2}{\pa^2\over \pa \p^2}\ln V\right ]\biggr \}
\label{scalar4}
\eea
The general procedure will be to study the eigenvalue problem for
 $\A$ (eq. \rf{bulk}) in the two regions where $V>0$ and $V<0$
separately,
and then impose suitable joining conditions on the partial derivatives
 of the eigenfunctions at the `turning point' $V=0$.

The simplest problem which we can analyze is that corresponding to
a  massless and minimally coupled scalar field\footnote{The case of a
conformally coupled scalar field turns out to be essentially equivalent.}
with
\beq
 {\cal{V}}(\p)=\l=0
\label{scalar5}
\eeq
 for which $V=-{a\over 2}$ and ${\cal{R}}=0$.
In this case, since the scale factor $a$ is {\it constrained} to be positive,
the evolution operator given by eq. \rf{scalar4} has no ambiguities due to
delta terms and becomes
\beq
\A=-{3\hbar^2\over 2a^2}\left [{\pa^2\over \pa a^2}+{1\over a}{\pa \over \pa a}
-{1\over a^2}{\pa^2\over \pa \p^2}\right ]
\label{scalar6}
\eeq


The hyperbolic problem described by eq. \rf{bulk} with the operator
 \rf{scalar6} and with $\E=1$ has been extensively discussed in the
literature
as an {\it ordinary differential problem} (see, e.g., \cite{Page} and
references
therein) or as a {\it Klein-Gordon eigenvalue problem} with respect to a
peculiar measure (see, e.g., \cite{Fulling}).
One can separate the differential equation and express bounded solutions
as a superposition of modified Bessel functions of imaginary order
\cite{Abramowicz}
\beq
\Psi_{\E, k}[a, \p]=\exp(\pm ik\p/\hbar)K_{ik\over 2\hbar}\left
({\sqrt{\E}a^2\over
\sqrt{6}\hbar} \right)
\label{scalar9}
\eeq
where $k$ is the separation constant,
or give them as a product of harmonic oscillator wave functions like
\beq
\Psi_{\E, n}[a, \p]=H_n\left [\left ({2\E \over 3\hbar^2}\right )^{1/4}X
\right ]H_n\left [\left ({2\E \over 3\hbar^2}\right )^{1/4}Y\right ]
e^{-\sqrt{\E \over 6}{(X^2+Y^2)\over \hbar}}
\label{scalar10}
\eeq
where $X=a\cosh \p$ and $Y=a\sinh \p$ \cite{Page}.

Unfortunately, neither of these solutions are orthogonal with respect
to the measure $\sqrt{\G}\propto a^3$, as it is easy to check.
To guarantee the self-adjointness of the evolution operator, one
has to construct a particular superposition of any of such solutions which
satisfies appropriate conditions on the boundary of the 2-d minisuperspace.

A possible method of solution of the
eigenvalue problem makes use of the so called `Rindler'
coordinates \cite{Rindler}.
Here we derive the formal boundary conditions for the eigenfunctions
required for Hermiticity, even though we have not been able to explicitly
solve for them.
The idea is to make the following transformation of coordinates in
the 2-d minisuperspace :
\bea
x&=&{a^2\over \sqrt{6}}\cosh 2\p
\non \\
y&=&{a^2\over \sqrt{6}}\sinh 2\p
\label{scalar11}
\eea
The minisuperspace line element in the new coordinates takes the simple form
 \hfil\break $ds_{Super}^2=(dy^2-dx^2)/4$.
The minisuperspace region $a\in [0, \infty)$ and $\p\in (-\infty, \infty)$
 is mapped into the right Rindler wedge of a 2-d Minkowski spacetime,
with boundary $\vert x\vert <y$.
The correspondence between old and new coordinates is one-to-one, with
$a=0$ mapped to $x=\pm y$, $\p=constant$ to ${x\over y}=constant$ lines
and $a=constant$ mapped to the hyperbolae $x^2-y^2=constant$.
The evolution operator simply becomes the 2-d wave operator in the Rindler
wedge, i.e.
\beq
\A=\hbar^2\left({\pa ^2\over \pa y^2}-{\pa ^2\over \pa x^2}\right)
\label{scalar12}
\eeq

By further `tilting' the minisuperspace axes by $45^o$ with the coordinates
\bea
x_+&=&{1\over 2}(x+y)
\non \\
x_-&=&{1\over 2}(x-y)
\label{scalar13}
\eea
($x_{\pm}\in [0, \infty)$), the evolution operator becomes
\beq
\A=-\hbar^2{\pa^2\over \pa x_+\pa x_-}
\label{scalar14}
\eeq
with constant measure.
The general solution of the differential equation \rf{bulk} is
thus a linear combination of plane waves
\beq
\Phi_{\E}[x_+, x_-]=\int d\lambda ~\left [A(\lambda)e^{i(\lambda x_+-{\E\over
\lambda}x_-)/\hbar}+
B(\lambda)e^{i(\lambda x_--{\E\over \lambda}x_+)/\hbar}\right]
\label{scalar21}
\eeq
Moreover, the self-adjointness condition for the evolution operator $\A$
requires that
\beq
\int_0^{\infty}dx_+~\int_0^{\infty}dx_-~\Phi_{\E_1}^{\ast}{\pa^2\over
\pa x_+\pa x_-}~\Phi_{\E_2}=\int_0^{\infty}dx_+~\int_0^{\infty}dx_-~\Phi_{\E_2}
{\pa^2\over \pa x_+\pa x_-}~\Phi_{\E_1}^{\ast}
\label{scalar22}
\eeq
Integrating by parts, one can see that this is true if the following
boundary condition for the eigenfunctions holds
\beq
\int_0^{\infty}dx_-\left [~\Phi_{\E_1}^{\ast}{\pa\over \pa x_-}~\Phi_{\E_2}
\right ]_{x_+=0}^{x_+=\infty}=\int_0^{\infty}dx_+\left [~\Phi_{\E_2}
{\pa\over \pa x_+}~\Phi_{\E_1}^{\ast}\right ]_{x_-=0}^{x_-=\infty}
\label{scalar23}
\eeq

Alternatively, one can transform the `infinite' boundary to a finite one
by use of the following conformal transformation on the superspace
 metric $ds_{Super}^2$:
\bea
d\tilde s_{Super}^2&=&\Omega^2ds_{Super}^2=-dudv
\non \\
u&=&2 ~\mbox{arctan}[(x-y)]~~~~,~~~~u\in [0, \pi]
\non \\
v&=&2 ~\mbox{arctan}[(x+y)]~~~~,~~~~v\in [0, \pi]
\non \\
 \Omega^2&=&16\cos^2\left({u\over 2}\right )\cos^2\left({v\over 2}\right)
\label{scalar24}
\eea
In this case the evolution operator transforms as
\beq
\A=-16\hbar^2\cos^2\left({u\over 2}\right)\cos^2\left({v\over 2}\right)
{\pa^2\over \pa u\pa v}
\eeq
with constant measure, and it is self-adjoint if
\beq
\int_0^{\pi}du\left [~\Phi_{\E_1}^{\ast}{\pa\over \pa u}~\Phi_{\E_2}
\right ]_{v=0}^{v=\pi}=\int_0^{\pi}dv\left [~\Phi_{\E_2}
{\pa\over \pa v}~\Phi_{\E_1}^{\ast}\right ]_{u=0}^{u=\pi}
\label{scalar25}
\eeq
The eigenfunctions are the same as those given in eq. \rf{scalar21},
provided one makes the substitutions $x_+\rightarrow 2\tan {v\over 2}$,
 $x_-\rightarrow 2\tan {u\over 2}$ and $\E\rightarrow \E/16$.

Thus, formally, the hyperbolic eigenvalue problem of eq. \rf{bulk}
for the 2-d minisuperspace model with the minimally coupled scalar
is solved provided one fixes the coefficients $A$ and $B$ in the
expansion \rf{scalar21} such that to satisfy given Cauchy
data and the condition \rf{scalar23} (or \rf{scalar25}) at the boundary.
In particular, although we cannot give an explicit formula,
the problem \rf{scalar25} certainly admits a solution.
A possible simple example would be, for instance, a wave function such that
$\Phi_{\E}(u, 0)$ ($\Phi_{\E}(0, v)$) has a fixed parity with respect to
the point $u={\pi \over 2}$ ($v={\pi \over 2}$).


\newpage

\noindent {\Large \bf Figure Captions}
\bigskip
\bigskip

\begin{description}
\item[Fig. 1] Classical evolution via $\partial_{\t}Q=\{Q,\A \}$.
The final configuration $q',q'',...$ depends on the $N,N_i$ functions,
but the classical action $S^\E[q_{f},q_0]$ from the initial to final
configurations is proportional to the evolution parameter $\t$,
independent of $N,N_i$.
\item[Fig. 2] Expansion of the (semiclassical) wavepacket.
WKB wavefunctions $\Psi[q,\t]$
are peaked, at any parameter time $\t$, on a set of
configurations $\{q_{cl}\}$ corresponding to hypersurfaces of the
classical 4-manifold, such that $\t=-S^\E[q_0,q_{cl}]/2\E$.
\item[Fig. 3] Evolution of a `triangular' wavepacket for a dust
dominated universe.
The parameters chosen are $n_0=400$, $a_0=1$, $\delta=0.02$ and
different times are labeled by $\t_{\ast}=-10^5\t$.
\item[Fig. 4] The modulus of the eigenfunction of $\A$ for the average
eigenvalue $\E_0=(3202)^2$ of the wave packet.
\item[Fig. 5] a) The wave packet at its closest approach to the
classical singularity.

b) The wave packet at the radius of maximum classical expansion.
\end{description}

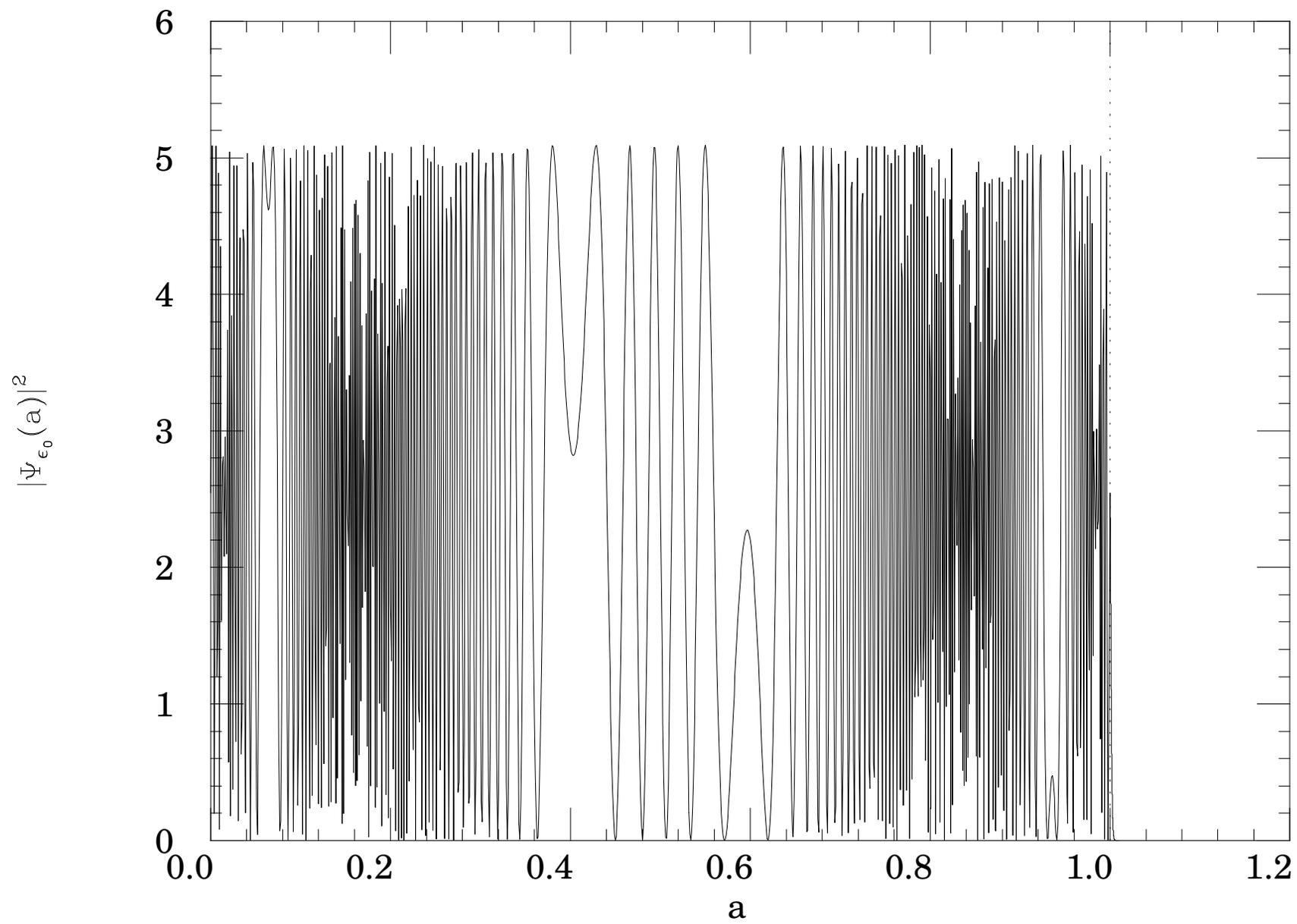
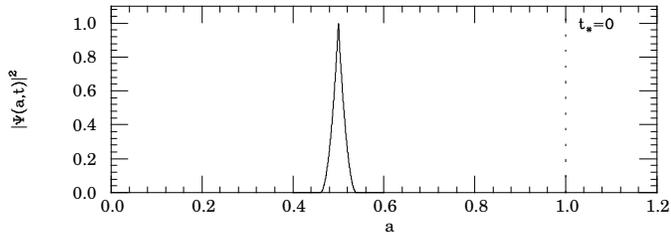
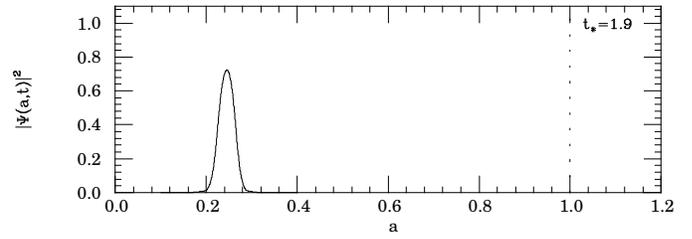
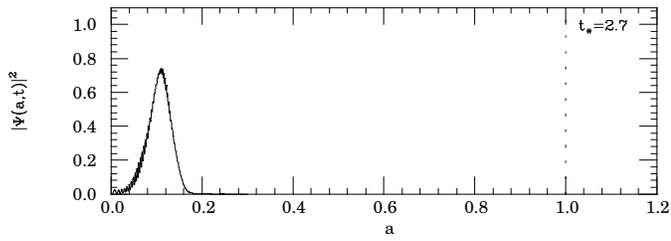
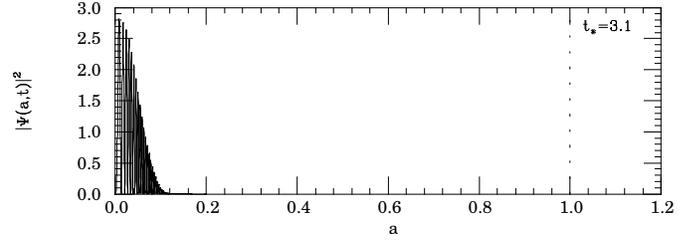
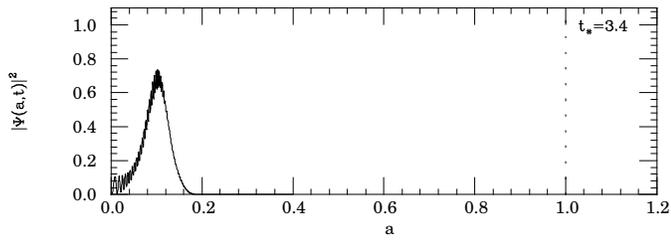
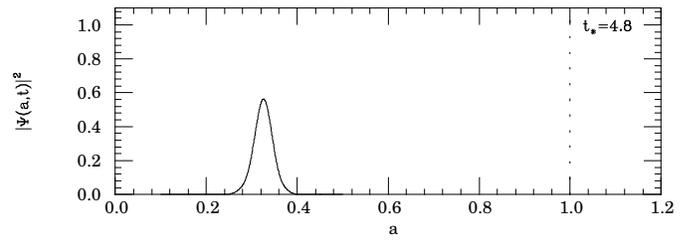
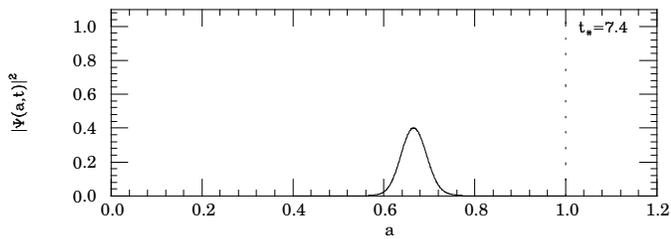
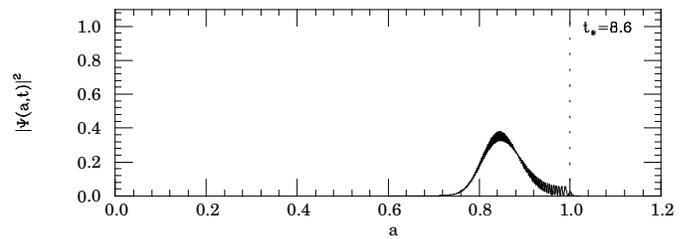
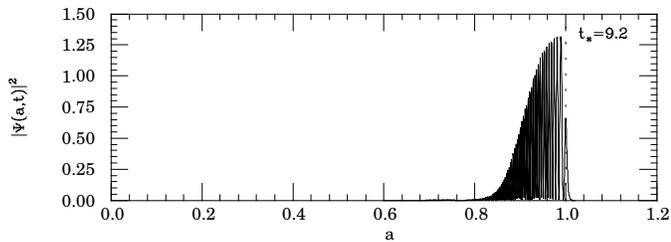
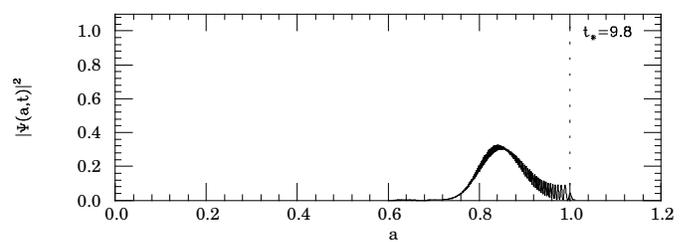
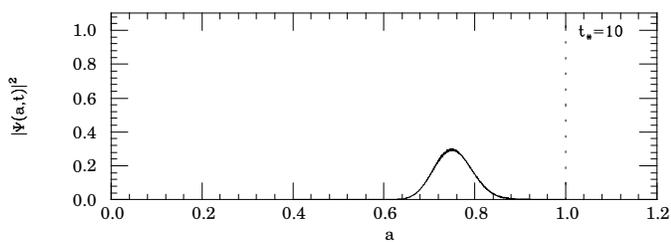
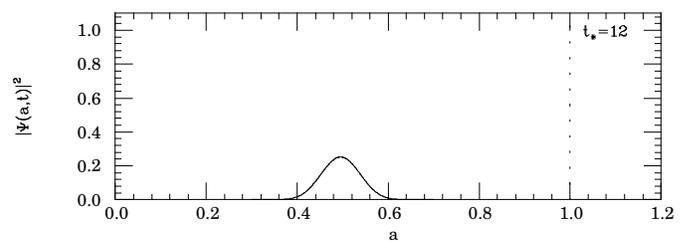
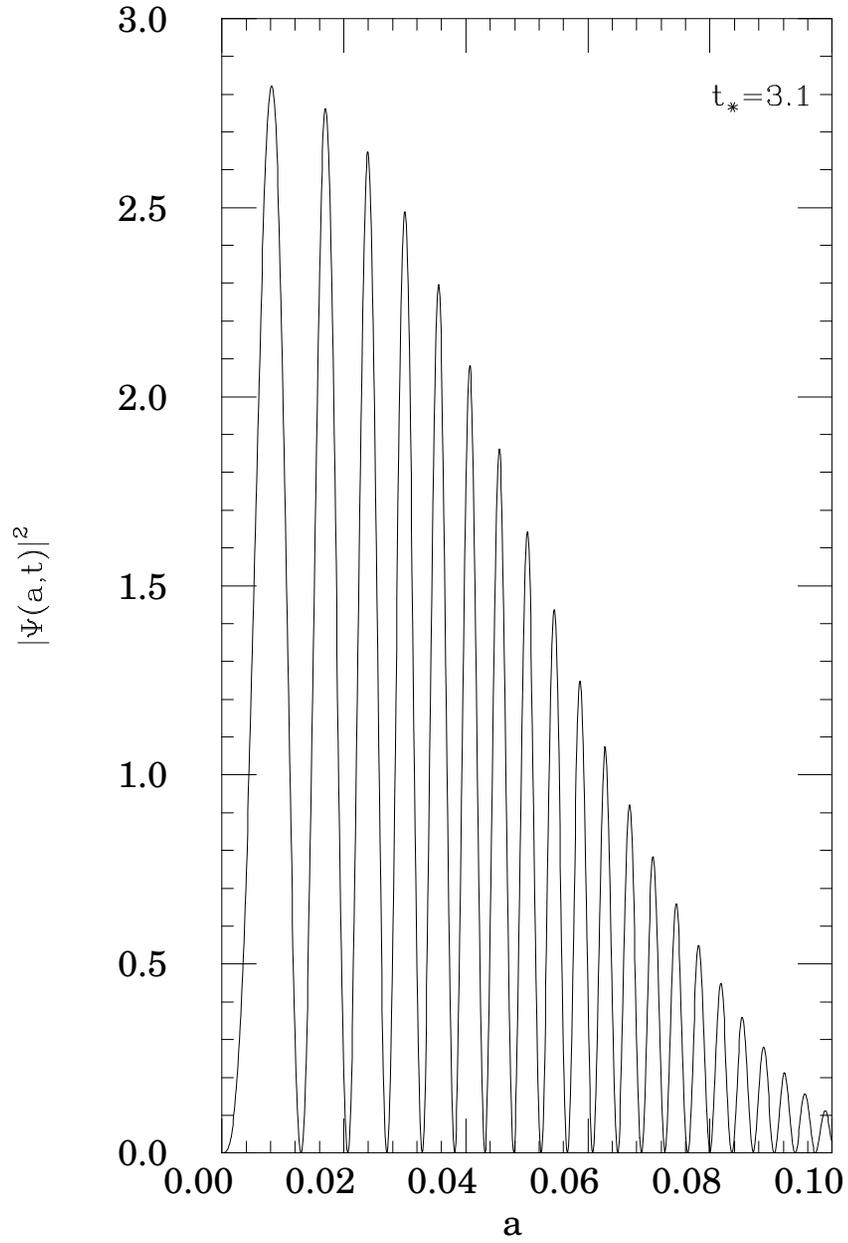
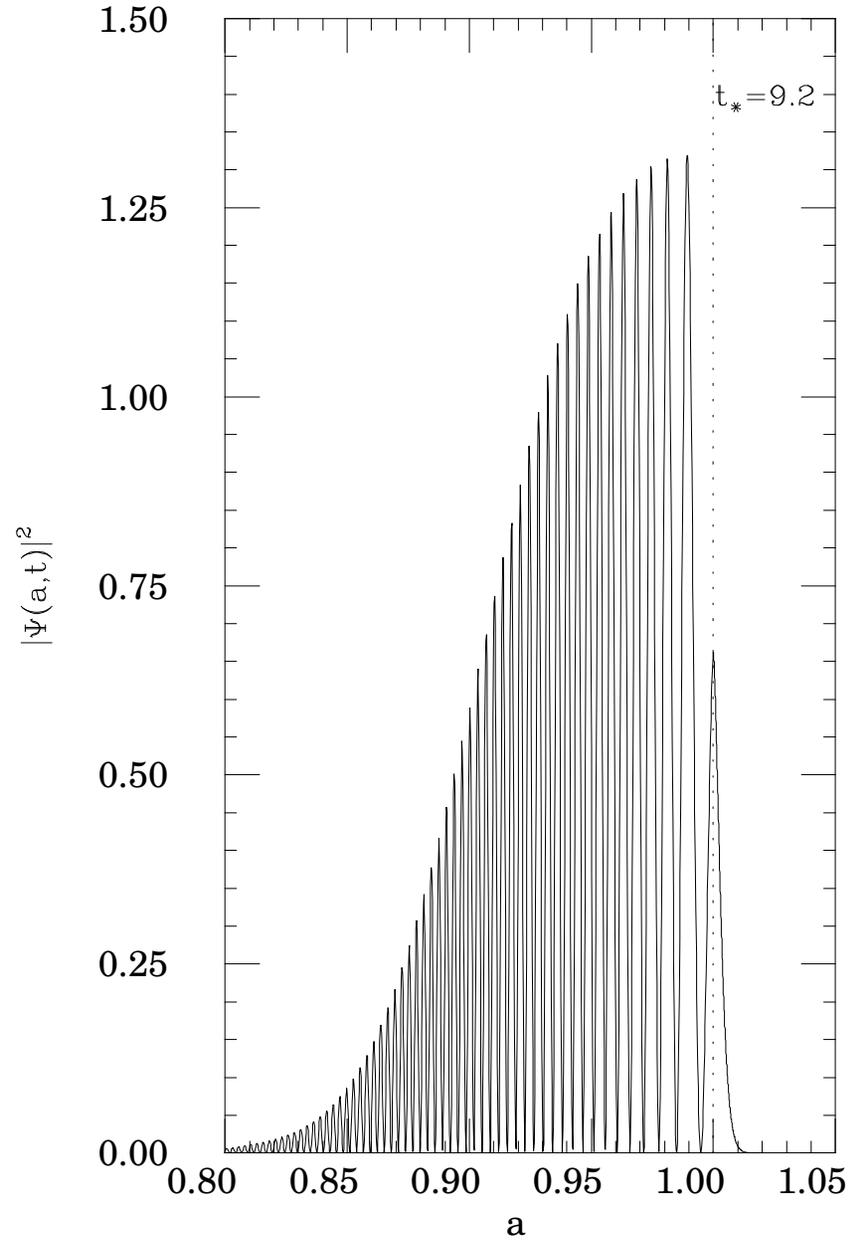
\end{document}